\documentclass[10pt, twocolumn]{iopart}
\usepackage{comment}
\usepackage{xcolor}
\usepackage{array}
\usepackage{soul}
\usepackage{graphicx} 
\usepackage{cite}
\usepackage{gensymb}
\usepackage[normalem]{ulem}
\usepackage{subcaption}
\expandafter\let\csname equation*\endcsname=\relax 
\expandafter\let\csname endequation*\endcsname=\relax 
\usepackage{amsmath}
\usepackage[margin=0.6in]{geometry}

\begin{document}

\twocolumn[
  \begin{@twocolumnfalse}
    \title{Power handling in a highly-radiative negative triangularity pilot plant}

    \author{M.A. Miller$^{1}$, D. Arnold$^{2}$, M. Wigram$^{1}$, A.O. Nelson$^{2}$, J. Witham$^{1}$,  G. Rutherford$^{1}$, H. Choudhury$^{2}$, C. Cummings$^{1}$, C. Paz-Soldan$^{2}$, D.G. Whyte$^{1}$}
    \date{April 2024}
    
    \address{$^{1}$~Plasma Science and Fusion Center, Massachusetts Institute of Technology, Cambridge, MA 02139, USA}
    \address{$^{2}$~Department of Applied Physics and Applied Mathematics, Columbia University, New York, NY 10027, USA}

    \maketitle
    
    \begin{abstract}
        This work explores detailed power handling solutions for a class of high-field, highly-radiative negative triangularity (NT) reactors based around the MANTA concept \cite{rutherford_manta_2024}. The divertor design is kept as simple as possible, opting for a standard divertor with standard leg length. FreeGS is used to create an equilibrium for the boundary region, prioritizing a short outer leg length of only $\sim$50 cm ($\sim$40\% of the minor radius). The UEDGE code package is used for the boundary plasma solution, to track plasma temperatures and fluxes to the divertor targets. It is found that for $P_\mathrm{SOL}$ = 25 MW and $n_\mathrm{sep} = 0.96 \times 10^{20}$ m$^{-3}$, conditions consistent with initial core transport modeling, little additional power mitigation is necessary. For external impurity injection of just 0.13\% Ne, the peak heat flux density at the more heavily loaded outer targets falls to 7.8 MW/m$^{2}$, while the electron temperature $T_\mathrm{e}$ remains just under 5 eV. Scans around the parameter space reveal that even at densities lower than in the primary operating scenario, $P_\mathrm{SOL}$ can be increased up to 50 MW, so long as a slightly higher fraction of extrinsic radiator is used. With less than 1\% neon (Ne) impurity content, the divertor still experiences less than 10 MW/m$^{2}$ at the outer target. Design of the plasma-facing components includes a close-fitting vacuum vessel with a tungsten inner surface as well as FLiBe-carrying cooling channels fashioned into the VV wall directly behind the divertor targets. For the seeded heat flux profile, Ansys Fluent heat transfer simulations estimate that the outer target temperature remains at just below 1550\degree C. Initial scoping of advanced divertor designs shows that for an X-divertor, detachment of the outer target becomes much simpler, and plasma fluxes to the targets drop considerably with only 0.01\% Ne content.
    \end{abstract}
  \end{@twocolumnfalse}
]

\section{Introduction}
\label{sec:Intro}

At the power density relevant to fusion power plants, divertors can be easily damaged by incident heat fluxes \cite{gunn_surface_2017}. It is expected that a tokamak operating in conditions necessary for high fusion gain will see unmitigated divertor parallel heat fluxes close to an order of magnitude greater than that of today's machines \cite{kuang_divertor_2020, menard_fusion_2022}. As a result, a great deal of research has been conducted to investigate advanced divertor configurations, including the X-divertor, the X-point target divertor, the snowflake divertor, and the super-X divertor, among others \cite{kotsch_xdiv_2004, covele_x-divertors_nodate, kotschenreuther_taming_2016}, all designed for improved heat flux handling. Regardless of divertor concept, any core scenario \emph{must} be compatible with a plasma edge that ensures divertor survivability. Rather than prioritizing core performance and hoping to find a solution that also enables adequate power handling, it is also possible to do the reverse and instead adopt a ``power-handling-first" reactor design philosophy \cite{kikuchi_l-mode-edge_2019}.

With this mantra in mind, this paper presents initial scoping and design of potential divertor configurations for fusion power plants similar to MANTA (Modular Adjustable Negative Triangularity ARC), a highly-radiative pilot plant concept that avoids edge-localized modes (ELMs) through the adoption of negative triangularity (NT) plasma shaping \cite{rutherford_manta_2024}. The adoption of NT scenarios creates conditions in the edge that are significantly more amenable to power handling than typical tokamak reactor concepts based around the positive triangularity H-mode. Given the ability for a negative triangularity reactor to maintain a high-density edge with low power fluxes crossing the separatrix, an advanced divertor design was not expected to be necessary. It is thought that MANTA's divertor, and NT divertors in general, can therefore feature relatively simple engineering, with short divertor legs and vertical target plates. 

\subsection{Advantages of negative triangularity}

\begin{figure}
\centering
\includegraphics[width=0.8\columnwidth]{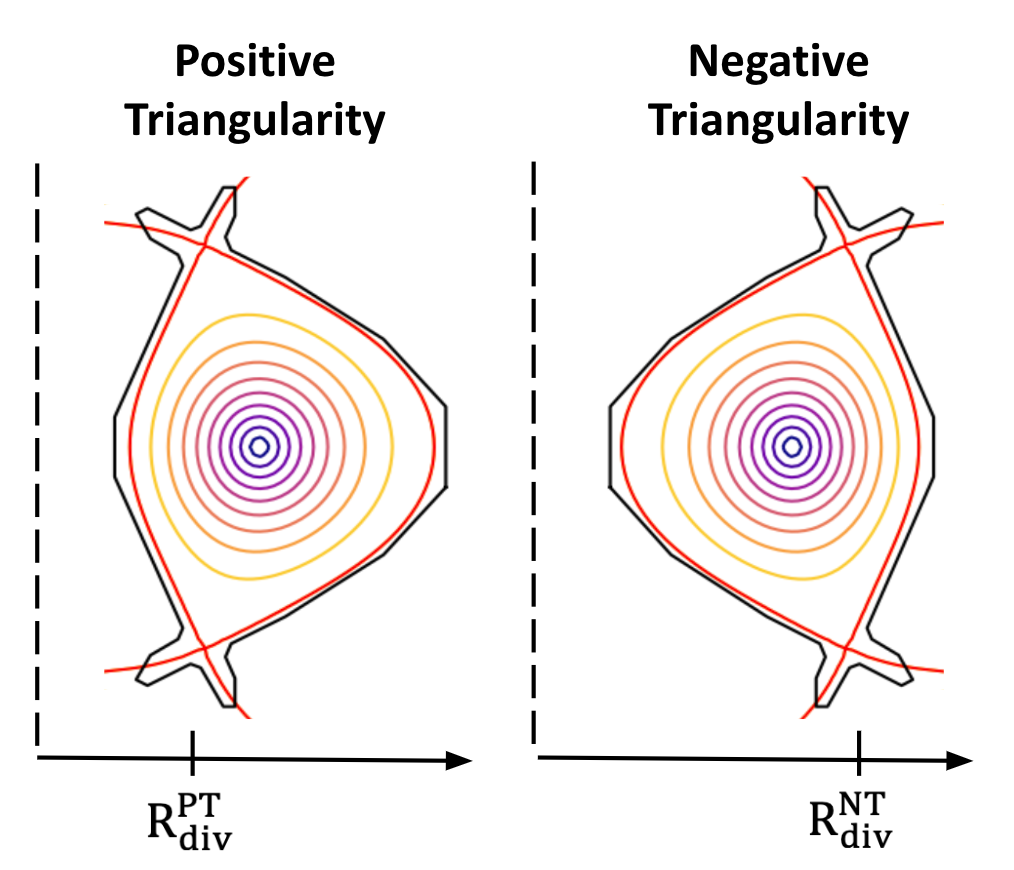}
\caption{Comparison of an example positive vs. negative triangularity plasma. Negative triangularity moves the divertor to a larger major radius.}
\label{fig:pt_vs_nt}
\end{figure}

NT research has grown in popularity significantly in recent years. Three tokamaks, TCV \cite{faitsch_dependence_2018-1, riva_shaping_2020, han_suppression_2021, Coda_2022}, DIII-D \cite{austin2019achievement, Marinoni_2019, marinoni_diverted_2021, nelson2022prospects, Nelson_2023, pazsoldan2023, thome_overview_2024}, and ASDEX Upgrade (AUG) \cite{happel_overview_2023} have successfully created NT-shaped plasmas. Many of these experiments have identified improvements in core performance. They also suggesting advantages for power handling when operating with a divertor designed for NT. The first of these is a geometric advantage in terms of a reduction in the heat flux density on the armored targets \cite{kikuchi_negative_2014}. NT moves the divertor to larger radii, as shown in Figure \ref{fig:pt_vs_nt}, thereby increasing its surface area compared to a positive triangularity (PT) plant of equivalent major radius.

Results for plasmas in NT suggest that it may be possible to sustain good core confinement even without an edge transport barrier, a scenario more closely resembling an ELM-free L-mode edge \cite{austin2019achievement, paz-soldan_plasma_2021, marinoni_diverted_2021, nelson2022prospects}. This is beneficial for power handling, as the absence of a pedestal and ELMs that accompany those in ELMy H-modes forgoes the need to survive large transients in heat and particle fluxes. Under the conditions expected in tokamaks at high plasma current, these will be particularly dangerous for machine survivability. Recent results on DIII-D bolster confidence in being able to retain this ELM-free edge for sufficiently negative triangularity, even at high heating powers \cite{Nelson_2023, nelson_characterization_2024}. It was also observed that this ELM-free operation was compatible with non-seeded detachment, even with a divertor not fully optimized for NT \cite{scotti_divertor_2024, scotti_high_2024}.

Other scenarios have also been proposed for reactors that emphasize the ``power-handling first approach". A recently proposed ARC-class reactor attempts to do this by remaining in L-mode while maintaining a high radiated core fraction, similarly to MANTA \cite{frank_radiative_2022}. Many other regimes appear attractive for reactor concepts, including the EDA H-mode \cite{greenwald_studies_2000}, the I-mode \cite{whyte_i-mode_2010}, the QH-mode \cite{burrell_quiescent_2002}, and the radiative X-point \cite{bernert_power_2017}. A radiative NT scenario, like that proposed in this work, offers similar ELM-free operation, with likely better confinement than the L-mode scenario and a likely broader heat flux width than the H/I-mode scenarios, as well as geometric advantages related to divertor location. Ability to access any of these regimes in a reactor remains uncertain, but if trends for benefits of NT extrapolate, these are perhaps easiest to access of all proposed regimes.

\subsection{Power handling in a highly radiative NT reactor}

Power handling in tokamaks can be divided into three categories: plasma power exhausted into the scrape-off layer (SOL), radiated power, and power deposited by neutrons. The power transported from the core into the SOL, $P_\mathrm{SOL}$, flows along open field lines to the divertor targets, potentially creating significant heat loads on the target plasma-facing components (PFCs). Thus, the magnitude of $P_\mathrm{SOL}$ strongly influences divertor complexity and lifetime. $P_\mathrm{SOL}$ depends on heat sources and sinks, and can be expressed as 
\begin{equation}
    P_\mathrm{SOL} = P_{\alpha} + P_\mathrm{oh} + P_\mathrm{aux} - P_\mathrm{rad} - dW/dt.
\end{equation} $P_{\alpha}$ comes from alpha particle heating, $P_\mathrm{oh}$ is the ohmic power generated by the tokamak's central solenoid, $P_\mathrm{aux}$ is any auxiliary power used to heat the plasma, $dW/dt$ is the time rate of change of the plasma stored energy, $W$, and $P_\mathrm{rad}$ is any power that is radiated away in the core, either by the main ion species or by impurities. 

Since NT FPPs like MANTA do not target H-mode operation, $P_\mathrm{SOL}$ need not exceed some threshold required for the transition. It is thought that H-mode access requires a certain critical level of ion heat flux to be driven through the edge. This is needed to build up a large enough ion pressure gradient to maintain $E \times B$ shear suppression and the development of a transport barrier in the typical H-mode edge \cite{burrell_effects_1997}. Without such a requirement, an NT reactor with a highly radiating core could, in principle, achieve a large range in fusion power, $P_\mathrm{fus}$, simply by varying core density and without concern for radiating too much power such that $P_\mathrm{SOL}$ falls below the L-H power threshold, $P_\mathrm{th}^\mathrm{L-H}$. Furthermore, since both $P_\mathrm{rad}$ and $P_\mathrm{fus}$ increase with $n$, as $n$ increases, $P_\mathrm{SOL}$ can remain low as long as $P_\mathrm{rad}$ can be controlled to compensate for changes in $P_\mathrm{fus}$. Details of MANTA's 0D analysis can be found in \cite{rutherford_manta_2024}.

The second exhaust channel is $P_\mathrm{rad}$. The highly-radiative core scenario considered here will radiate power throughout the plasma volume, including the core, common flux region (CFR), and divertor. The current analysis of the operating point does assume that the radiated fraction is a controllable parameter in NT plasmas, as it sets $P_\mathrm{SOL}$. The calculation of core radiation figures into transport analysis explained in more detail in \cite{wilson_integrated_2024}, and the assertion of controllable radiation fraction is an active area of investigation. Radiation in the CFR and divertor is considered here, and it is included in the transport calculation described in Section \ref{sec:uedge}. Success of a power handling solution requires both sufficiently low incident heat flux on the divertor plates to prevent damage as well as survivable radiation loading on the main chamber.

The final power exhaust channel is that of the 14.1 MeV neutrons, $P_{n}$, released via fusion reactions. While these neutrons do not impart substantial heat flux onto MANTA's PFCs, avoiding substantial neutron flux to achieve reasonable magnet lifetime presents a strong constraint on the locations of the poloidal field (PF) coils. Thus, $P_{n}$ indirectly affects the achievable divertor configurations. The PF coil set used throughout this work balances the choice of long lifetimes or low currents, as detailed in Section \ref{subsec:equ}.

While trends identified in this work are representative of a MANTA-like concept, the operating point scoped in this paper is that of an earlier iteration of MANTA, rather than the final MANTA design point \cite{rutherford_manta_2024}. However, this is still a self-consistent design and allows for representative scoping of divertor operation for negative triangularity tokamaks similar to MANTA. The remainder of the paper is organized as follows. Section \ref{sec:manta_recap} provides a recap of the design of MANTA from its 0D scoping, to the generation of its equilibrium using FreeGS. Section \ref{sec:uedge} details the SOL simulation workflow using the UEDGE code. Section \ref{sec:operational_space} tests the robustness of the divertor solution to deviations from assumptions in the edge model. Sections \ref{sec:vv_design} and \ref{sec:vv_results} discuss how the outputs of the UEDGE solution have been utilized to design and determine power loading on the structural components of MANTA's divertor. Section \ref{sec:adv_div} presents initial scoping of advanced divertor concepts, namely the X-divertor. Finally, Section \ref{sec:conclusions} will summarize the findings of the scoping work presented here.

\section{The baseline MANTA design}
\label{sec:manta_recap}

The following metrics represent 0D heuristics to quantify the divertor challenge of a particular device. Higher values represent increased power handling difficulty.
\begin{align}
    M_1 &= P_\mathrm{SOL}B_{T}/{R_0}\\
    M_2 &= (P_\mathrm{SOL}B_{T}/R_0)/n_\mathrm{sep}^{2}
\end{align}
Here, $B_{T}$ is the toroidal magnetic field, $R_0$ is the major radius, and $n_\mathrm{sep}$ is the density at the separatrix. Equation (2) characterizes the parallel heat flux density, $q_{||} \propto P_\mathrm{SOL}B_{T}/{R}$ ($\lambda_q \propto B_{p}$ \cite{Eich_Hmode_2013}, where $\lambda_q$ is the SOL heat flux width, computed as the fall-off length of $q_{||}$). By including the plasma density, Expression (3) adds dissipation physics to the metric (since power dissipation scales $\propto n_\mathrm{sep}^{2}$), following the Lengyel model \cite{Moulton_Lengyel_2021}). For a variety of next-generation proposed devices as well as reactor-class designs, these metrics are considerably higher than MANTA's. Their values are listed in Table \ref{tab:divertorMetrics}.

\begin{table*}[hbtp!]
\begin{center}
\caption{Comparison of divertor metrics between MANTA \cite{rutherford_manta_2024}, ARC V1 \cite{Sorbom_APS_2020}, ITER \cite{pitts_physics_2019}, EU-DEMO \cite{Reimerdes_DEMO_2020}, Japan-DEMO \cite{asakura_studies_2017}, and CFETR \cite{CFETR_optimization}}
\label{tab:divertorMetrics}
\begin{tabular}{lllllll}
 \hline
 Parameter & MANTA & ARC V1c & ITER & CFETR & EU-DEMO & Japan-DEMO \\
 \hline
 \hline
 $P_\mathrm{fus}$ (MW) & 451 & 500 & 500 & 558 & 2000 & 1694 \\
 $R$ (m) & 4.55 & 3.65 & 6.2 & 7.2 & 8.8 & 8.5  \\
 $B_T$ (T) & 11.1 & 11.6 & 5.3 & 6.5 & 5.8 & 5.94  \\
 $P_\mathrm{SOL}$ (MW) & 23.5 & 83 & 100 & 91 & 150 & 258 \\
 $n_\mathrm{sep} (10^{20} m^{-3})$ & 0.9 & 0.61* & 0.45 & 0.25 & 0.25 & 0.2 \\
 $P_\mathrm{SOL}B_{T}/{R}$ & 57.3 & 263 & 85.5 & 82.2 & 98.9 & 369 \\
 $(P_\mathrm{SOL}B_{T}/R)/n_\mathrm{sep}^{2}$ & 70.7 & 707 & 422 & 1310 & 1580 & 9230 \\
 \hline
\end{tabular}
\end{center}
\end{table*}

Among all future devices listed, MANTA has the lowest value of $M_{1}$. This result comes from MANTA's ability to operate at high radiated fraction and therefore, low $P_\mathrm{SOL}$. $M_{1}$ for MANTA, however, is only slightly lower than for ITER and CFETR, mostly as a result of their $>50\%$ larger $R_0$. In particular, MANTA only has $\sim$30\% lower $M_{1}$ than CFETR, another reactor-class device. EU-DEMO, however, at a larger $P_\mathrm{SOL}$ than ITER, has $\sim$60\% higher $M_{1}$ than MANTA. MANTA's power handling task is close to four times as easy as ARC's, another high field device, which must handle much higher $P_\mathrm{SOL}$ values in order to stay in H-mode, and which can be variable and depend on the produced fusion power $P_\mathrm{fus}$. $M_{1}$ may even be slightly conservative, given that for a given $R_{0}$, an NT divertor is at even larger $R$ than a PT divertor, due to its shape. 

The advantage of an NT reactor's ability to operate at high $n_\mathrm{sep}$ becomes clear when considering $M_{2}$. The NT edge features shallow density gradients, meaning for a given core density, MANTA can support a high $n_\mathrm{sep}$ while maintaining high fusion performance \cite{nelson_characterization_2024, rutherford_manta_2024}. For MANTA, $M_{2}$ = 70.7, while the other devices operating in H-mode have much larger values. These designs are at considerably lower $n_\mathrm{sep}$, making $M_{2} >$ 1000 for CFETR, EU-DEMO, and Japan-DEMO. ARC, which plans to operate at slightly higher $n_\mathrm{sep}$ than other reactors, still has a value for $M_2$ that is 10 times larger than that for MANTA. For reference, some of the key design parameters of this preliminary MANTA operating point are listed in Table \ref{tab:MANTA_params}. See\cite{rutherford_manta_2024} for the final values of these and additional parameters.

\begin{table}[h!]
\begin{center}
\caption{Additional MANTA key design parameters}
\label{tab:MANTA_params}
\resizebox{\columnwidth}{!}{
\begin{tabular}{lll}
 \hline
 Parameter & Symbol & Value\\
 \hline
 \hline
Total thermal power & $P_{\rm th}$ & 530 MW\\
Net electric power & $P_{\rm e,net}$ & 90 MWe\\
ICRF coupled power & $P_{\rm ICRH}$ & 40 MW\\
Plasma gain & $Q$ & 11.5\\
Electricity gain & $Q_E$ & 2.4\\
Major radius & $R_0$ & 4.55 m\\
Plasma minor radius & a & 1.2 m\\
Plasma elongation & $\kappa$ & 1.4\\
Plasma triangularity & $\delta$ & -0.5\\
Plasma current & $I_P$ & 10 MA\\
Greenwald fraction & $f_{G}$ & 0.88\\
Safety factor at $\Psi_{N}$ = 0.95 & $q_{95}$ & 2.3\\
Minimum safety factor & $q_{\rm min}$ & 0.905\\
Energy confinement time & $\tau_{E}$ & 0.94 s\\
H$_{89}$ confinement factor & H$_{89}$ & 1.44\\
 \hline
\end{tabular}
}
\end{center}
\end{table}

\subsection{Poloidal field coil location and current optimization}
\label{subsec:equ}

\begin{figure}
\centering
\includegraphics[width=\columnwidth]{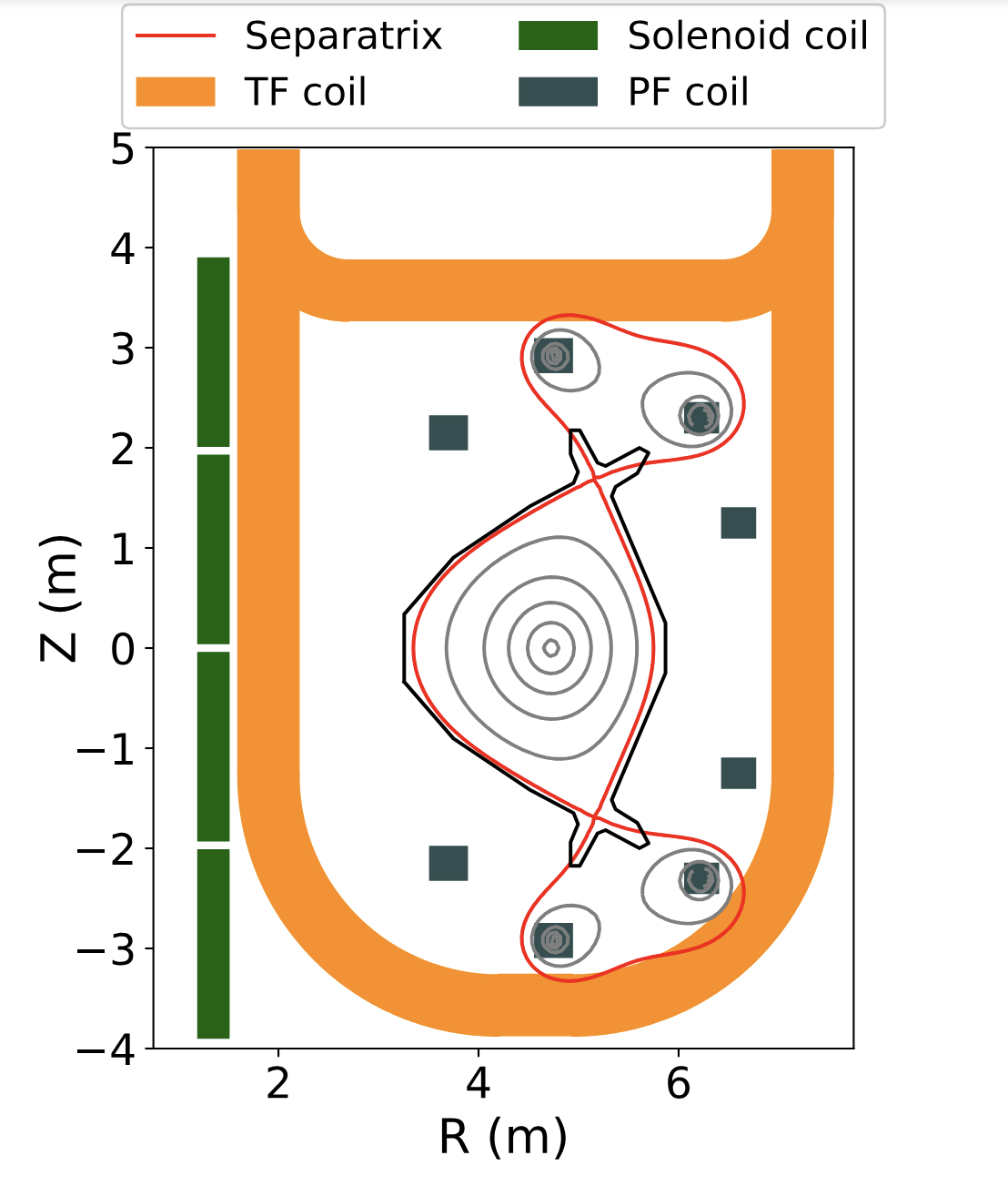}
\caption{The poloidal cross-section used in this work, which is modified slightly from the MANTA design presented in reference \cite{rutherford_manta_2024}.}
\label{fig:freeGS_equi}
\end{figure}

The NT magnetic equilibrium used in this work differs from the final MANTA equilibrium in that a fourth poloidal field (PF) coil was employed. Beginning with an equilibrium generated by the Grad-Shafranov solver CHEASE\cite{LutjensCHEASE} based off an early 0D scoping, the divertor geometry and PF coil locations/currents were developed with another Grad-Shafranov solver, FreeGS\cite{FreeGS}, which is able to model the X-point and divertor region. The resulting equilibrium is shown in Fig. \ref{fig:freeGS_equi}, where parabolic fits were used for the current and pressure profiles. The over-arching goal while designing the PF coilset was to keep the divertor as simple as possible, meaning no complex magnetic topologies and minimal X-points, while still surviving $P_\mathrm{SOL} = 25$ MW. This resulted in four pairs of PF coils, which were placed inside the toroidal field (TF) coils to reduce their cost. This was possible because MANTA's TF coils are demountable, allowing access to their interior. The PF coil locations shown in Figure \ref{fig:freeGS_equi} were determined through moving the coils by hand and running neutronics simulations for each new coilset until the PF coil lifetimes were deemed acceptable from the perspective of reactor cost. The resulting PF locations and currents are given in Table \ref{tab:PF_params}.

\begin{table}
\begin{center}
\caption{Poloidal field coil parameters}
\label{tab:PF_params}
\resizebox{\columnwidth}{!}{
\begin{tabular}{lllll}
 \hline
 Parameter & PF1 & PF2 & PF3 & PF4\\
 \hline
 \hline
Current$\cdot$turns [MA$\cdot$turns] & -7.8 & 6.1 & 2.1 & -4.3\\
Number of turns & 86 & 67 & 24 & 48\\
Height [m] & 0.33 & 0.29 & 0.14 & 0.29 \\
Width [m] & 0.37 & 0.33 & 0.22 & 0.22\\
R [m] & 3.65 & 5.05 & 6.25 & 6.65 \\
Z [m] & $\pm$2.25 & $\pm$2.9 & $\pm$2.5 & $\pm$ 1.25 \\
 \hline
\end{tabular}}
\end{center}
\end{table}

\section{UEDGE simulations of scrape-off layer (SOL) plasma}
\label{sec:uedge}

From this equilibrium, UEDGE was employed to calculate the heat and particle transport through the MANTA SOL to the divertor target plates. UEDGE solves the 2D Braginskii plasma fluid equations for magnetized plasmas in the tokamak edge. UEDGE also includes a fluid neutral solver that models the particle source and neutral dynamics in the simulation domain \cite{rognlien_fully_1992}. The code uses an implicit numerical scheme to advance the plasma solution in time. UEDGE has been previously used as the workhorse for scoping divertor parameter space and assessing divertor survivability in other high-field reactor-class devices \cite{ballinger_dependence_2022}. To ensure divertor survivability, two key metrics are tracked. The first and most important is the maximum perpendicular heat flux density on the targets, $q_\mathrm{\perp,surf}^\mathrm{max}$, simply referred to as $q_\mathrm{surf}$ hereafter. The second is temperature of ions and electrons, $T_{i}$ and $T_{e}$ respectively, on the target surface. These two metrics, taken together, must be met simultaneously to respect limits of recrystallization of the divertor's tungsten surface \cite{pitts_physics_2019}.

\begin{figure}
\centering
\includegraphics[width=\columnwidth]{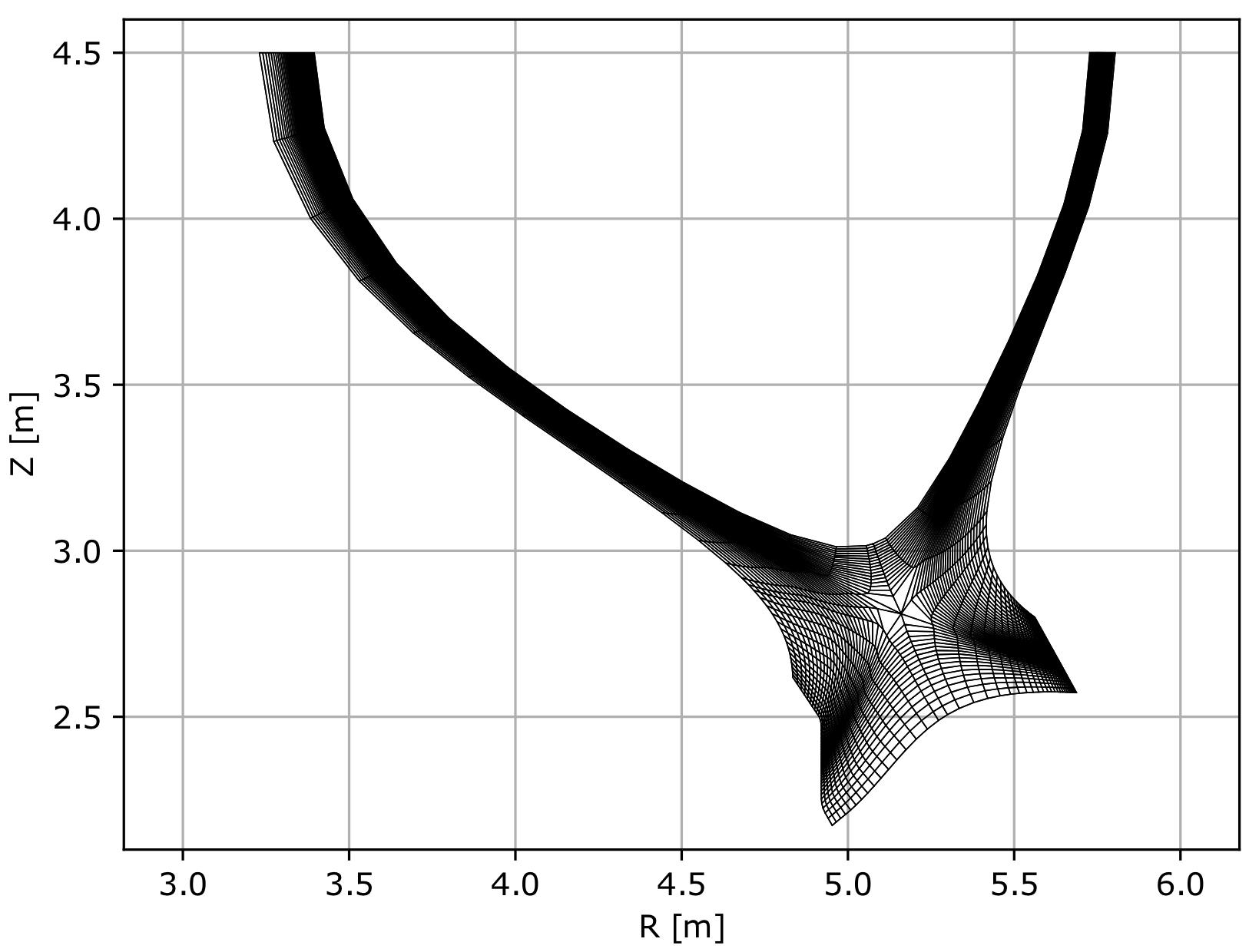}
\caption{Computational domain for UEDGE simulations generated from FreeGS equilibrium shown in Figure \ref{fig:freeGS_equi}.}
\label{fig:uedge_mesh}
\end{figure}

The in-built UEDGE mesh generator was used to create the grid shown in Figure \ref{fig:uedge_mesh}, focused on high resolution near the separatrix and near the targets, important for resolving steep plasma gradients and physics of plasma-wall interactions in these regions. The resolution was chosen to be on the order of the parallel heat flux gradient scale length expected near the separatrix at the outer midplane (OMP), as explained in sub-section \ref{subsec:transport}. MANTA plans to operate in a double-null (DN) magnetic topology, doubling the plasma wetted area on which to divert plasma exhaust. For simplicity and to increase computational efficiency of the simulations, up-down symmetry is assumed, and only the bottom half of the SOL was simulated. A reflection condition is imposed at midplane poloidal boundary. Such a setup is not consistent with drift transport terms, and so drifts are not included. Near the divertor targets, a non-orthogonal grid is used, which allows for precise target geometry tuning relative to the field lines imposed by divertor PF coils calculated with FreeGS above.  The outer and inner targets are poloidally tilted such that field line grazing angles at the target plates are 5.8\textdegree\ and 2.5\textdegree, respectively. Further tilting to reduce the grazing angles to $\sim$2\textdegree\, would potentially provide additional power handling capability, but mesh cell distortion set a numerical limit that prevented this. UEDGE utilizes an implicit numerical scheme to iterate the plasma solution until steady-state is reached, by slowly increasing the simulation timestep until residuals have dropped below a specified tolerance of 10$^{-10}$.

\subsection{Heat and particle sources}
\label{subsec:sources}
Core transport modeling for MANTA found a fixed $P_\mathrm{SOL}$ = 25 MW to be compatible with a fusion core that satisfied the requirements of the NASEM report. The L-H transition power from the Martin scaling \cite{martin_power_2008} for MANTA is $P_\mathrm{th}^\mathrm{L-H} \approx 100$ MW, meaning MANTA will operate at $f_\mathrm{L-H} = P_\mathrm{SOL}/P_\mathrm{th}^\mathrm{L-H}$ = 0.25. This value of $P_\mathrm{SOL}$ is thus used for the UEDGE point design. All 25 MW are applied at the core boundary, a few mm inside of the separatrix, and the power is assumed to be divided evenly between electrons and ions. The difference in power a few mm inside of the separatrix and at the separatrix is not expected to represent a significant fraction of $P_\mathrm{SOL}$. The exact partition of energy between electrons and ions, however, is a source of uncertainty, but is out of the scope of this paper. For the density boundary condition in MANTA's design, we use the result of an integrated modeling workflow described in \cite{rutherford_manta_2024}. This determines a value of $n_\mathrm{sep} = 9.6 \times 10^{19}$ m$^{-3}$ yielded an edge solution compatible with that of the core. Particle balance on MANTA is achieved through cryopumping, with the objective of removing He ash produced in the core. This surface is designated at the PFR boundary on the outboard side, near the outer target where the highest neutral pressure and thus, pumping efficiency, is expected. A pump removal rate of $\sim$5$\times 10^{21}$s$^{-1}$ is chosen to meet $^{4}$He removal requirements, assuming a 2\% $^{4}$He  concentration in the plasma exhaust from the reactor core and an enrichment of $\sim$0.75 in the divertor like that of MANTA \cite{rutherford_manta_2024} (consistent with previous studies \cite{Hillis_enrichment_1999}).

\subsection{Transport model}
\label{subsec:transport}

AS UEDGE does not resolve turbulent temporal and spatial scales, radial fluxes are calculated using a set of user-supplied cross-field transport coefficients. Since there is a lack of radiative L-mode negative triangularity experiments to compare to and current physics models for edge transport are incomplete, it is necessary to use existing scalings to choose a transport model. To do so, three main quantities are used to guide transport model selection: the e-folding length of the parallel heat flux at the separatrix, $\lambda_{q}$, the e-folding length of the plasma density at the separatrix, $\lambda_{n}$, and the ratio of inboard to outboard power, $P_\mathrm{in-out}$. Predictions for these are made for their values using empirical scalings and data from high-field devices.

\begin{figure}[h!]
\centering
\includegraphics[width=1\columnwidth]{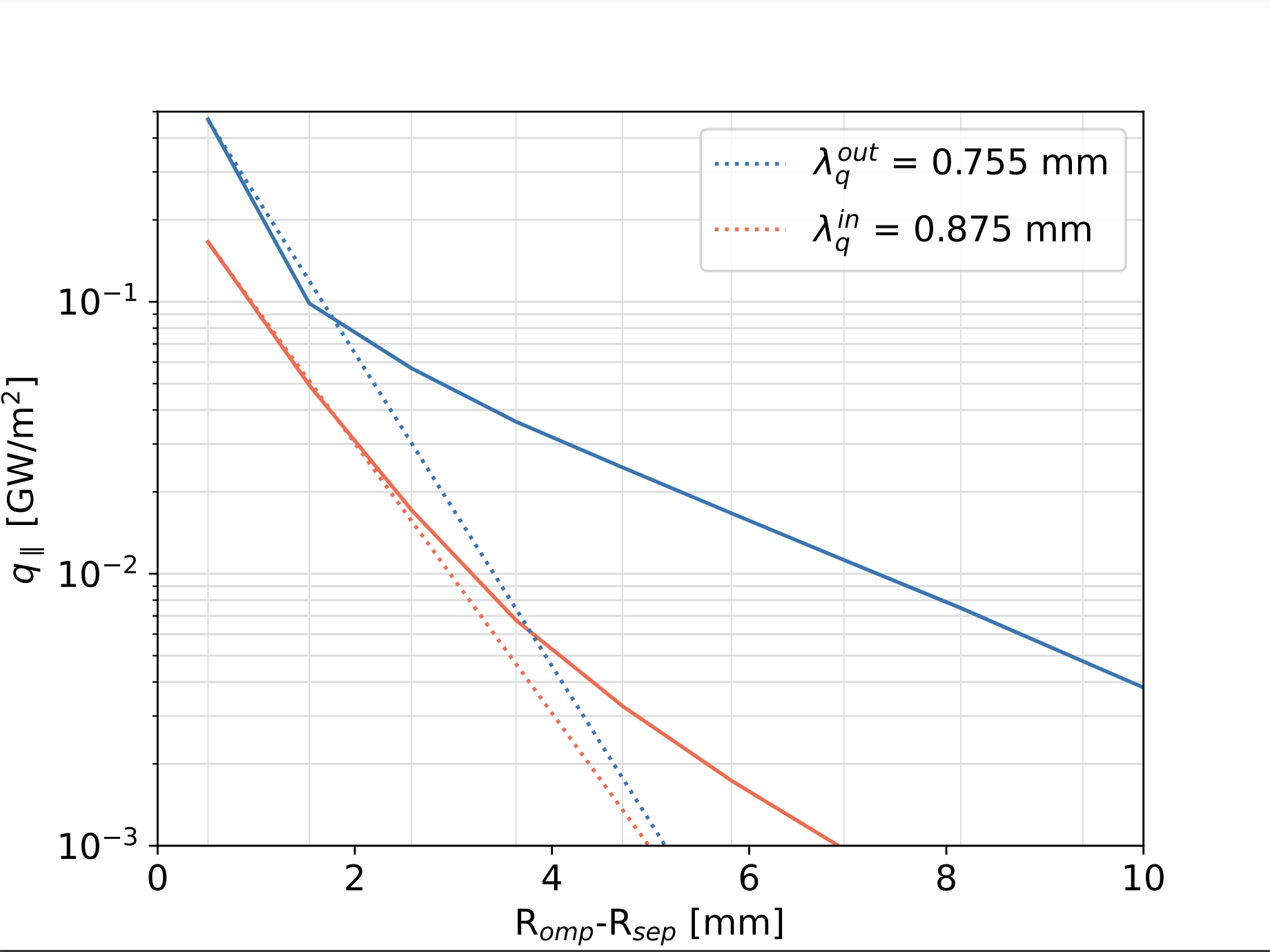}
\caption{Parallel heat flux entering the divertor, calculated at the poloidal cut corresponding to the X-point. Radial coordinate is the distance from the separatrix mapped to the midplane. $q_{\parallel}$ is calculated both on the inboard side (orange) and outboard side (blue). $\lambda_{q}$ is taken at this poloidal location, using midplane coordinates, and chosen to be $\sim$0.75 mm on the outboard side.}
\label{fig:xpt_lambdaq}
\end{figure}

The first of these, $\lambda_{q}$, is very influential for power handling. It dictates the size of the divertor wetted area that must handle the bulk of the power crossing the separatrix. The decay length in the near-SOL for the heat flux in an NT edge is likely to be similar to that in H-mode, though perhaps not as small. A variety of scalings for $\lambda_{q}$ scalings\cite{eich_empiricial_2013, brunner_high-resolution_2018, horacek_scaling_2020} serve to provide bounds that MANTA's $\lambda_{q}$ is apt to fall in. The results for these scalings for MANTA's primary operating parameter set are tabulated in Table \ref{tab:lambdaq_scalings}. Interestingly enough (and perhaps unfortunately), Eich and Brunner's scalings both predict heat flux widths possibly as low as 0.3 - 0.4 mm. Eich's scaling is a unified scaling, and succeeds in predicting $\lambda_{q}$ in many current devices spanning a large range of operational paremters. It is, however, purely an H-mode scaling. Brunner's on the other hand is multi-regime and might fare well in an NT edge, but it was constructed solely on data from Alcator C-Mod. Other scalings of C-Mod data by Brunner, but using $B_{p}$ could suggest $\lambda_{q}$ three times as high if the NT edge is to resemble the L-mode. Further, a multi-machine scaling by Horacek would put MANTA's $\lambda_{q} >$ 10 mm. Considering this variation in possible values of $\lambda_{q}$ and in the interest of computational efficiency associated with grid size resolution, a middling value of $\lambda_{q} \approx$ 0.75 mm  at the divertor throat is chosen. Figure \ref{fig:xpt_lambdaq} shows the parallel heat flux profiles at the X-point, with fits to the near-SOL $q_{\parallel}$, from which $\lambda_{q}$ is extracted.

\begin{table*}[hbtp!]
\begin{center}
\caption{Values of $\lambda_{q}$ for different scalings}
\label{tab:lambdaq_scalings}
\begin{tabular}{llll}
 \hline
 Scaling parameter & Confinement mode & Device & Predicted $\lambda_{q} (mm)$ \\
 \hline
 \hline
 $B_{p}^{-1.19}$\cite{eich_empiricial_2013} & H & multi-machine & 0.39 \\
 $<p>^{-0.48}$\cite{brunner_high-resolution_2018} & all & C-Mod & 0.30 \\
 $B_{p}^{-0.74}$\cite{brunner_high-resolution_2018} & L & C-Mod & 1.01 \\
 $B_{p}^{-0.57}$\cite{brunner_high-resolution_2018} & I & C-Mod & 0.75 \\
 $B_{p}^{-0.96}$\cite{brunner_high-resolution_2018} & H & C-Mod & 0.61 \\
 $B_{p}^{-0.36}q_{95}^{0.55}f_{GW}^{0.92}$\cite{horacek_scaling_2020} & L & multi-machine & 11.82 \\
 \hline
\end{tabular}
\end{center}
\end{table*}

For $\lambda_{n}$ and $P_\mathrm{in-out}$, there is even less predictive capability than for $\lambda_{q}$ and no empirical scalings to extrapolate from. Instead, these are chosen by considering measurements from Thomson Scattering and Langmuir probes on current experiments \cite{silvagni_scrape-off_2020, ballinger_dependence_2022}. Using these data, it is expected that $\lambda_{n} = \frac{n}{\nabla n}$ at MANTA's high $B_{p}$ may be somewhere around 6 - 10 mm near the separatrix for L-mode plasmas. Finally, to determine $P_\mathrm{in-out}$, data for power flow to the inner and outer divertor of current devices is used. Ballooning-like transport on the LFS is thought to carry significantly more particles to the outer divertor than the inner divertor \cite{labombard_particle_2001, DeTemmerman_asymmetries_2010, Brunner_powersharing_2018}. This effect is due in part to a large volume of ``bad curvature" on the outboard side, making it more susceptible to interchange-driven turbulence \cite{lim_turbulence_2023}. Because of the flipped cross-section, this type of turbulence may not be as detrimental in NT because of the decreased surface area of the bad-curvature LFS separatrix. On the other hand, while DN configurations enable power-sharing between the top and bottom targets, the inner and outer SOLs are disconnected, leading to a larger in-out power asymmetry than witnessed in single null (SN) cases. Thus, MANTA is assumed to have a 30-70 power split between the inner and outer divertor targets, respectively, corresponding to $P_\mathrm{in-out} = P_\mathrm{in}/(P_\mathrm{in} + P_\mathrm{out}) = 0.3$. 

The primary actuators used to achieve these values for $\lambda_{q}$, $\lambda_{n}$, and $P_\mathrm{in-out}$ are the particle and electron/ion heat diffusivities, $D$ and $\chi_{e,i}$ respectively. As a result of difficulty in measuring the ion temperature, it is hard to constrain $T_{i}$, although it is known that ion to electron temperature ratio, $\tau_{i} = T_{i}/T_{e}$ can be anywhere between 2 - 6 in the edge \cite{brunner_assessment_2013}. For simplicity, $\chi_{i}$ is chosen equal to $\chi_{e}$, a common assumption made in the SOL \cite{ballinger_simulation_2021}. Further, these transport coefficients are applied uniformly in the radial direction, yielding an entirely flat transport profile without a ``transport well" characteristic of an H-mode pedestal. Modeling with radially invariant transport profiles in the open field line region is seen to reproduce gradients typical of L-mode confinement for the near-SOL. Though this may miss some features of the far-SOL (e.g. density shoulders, filament transport, etc), the focus of this study is reproducing the expected near-SOL heat flux width properties and power exhaust, so these far-SOL features are not considered.

\begin{figure}
\centering
\includegraphics[width=\columnwidth]{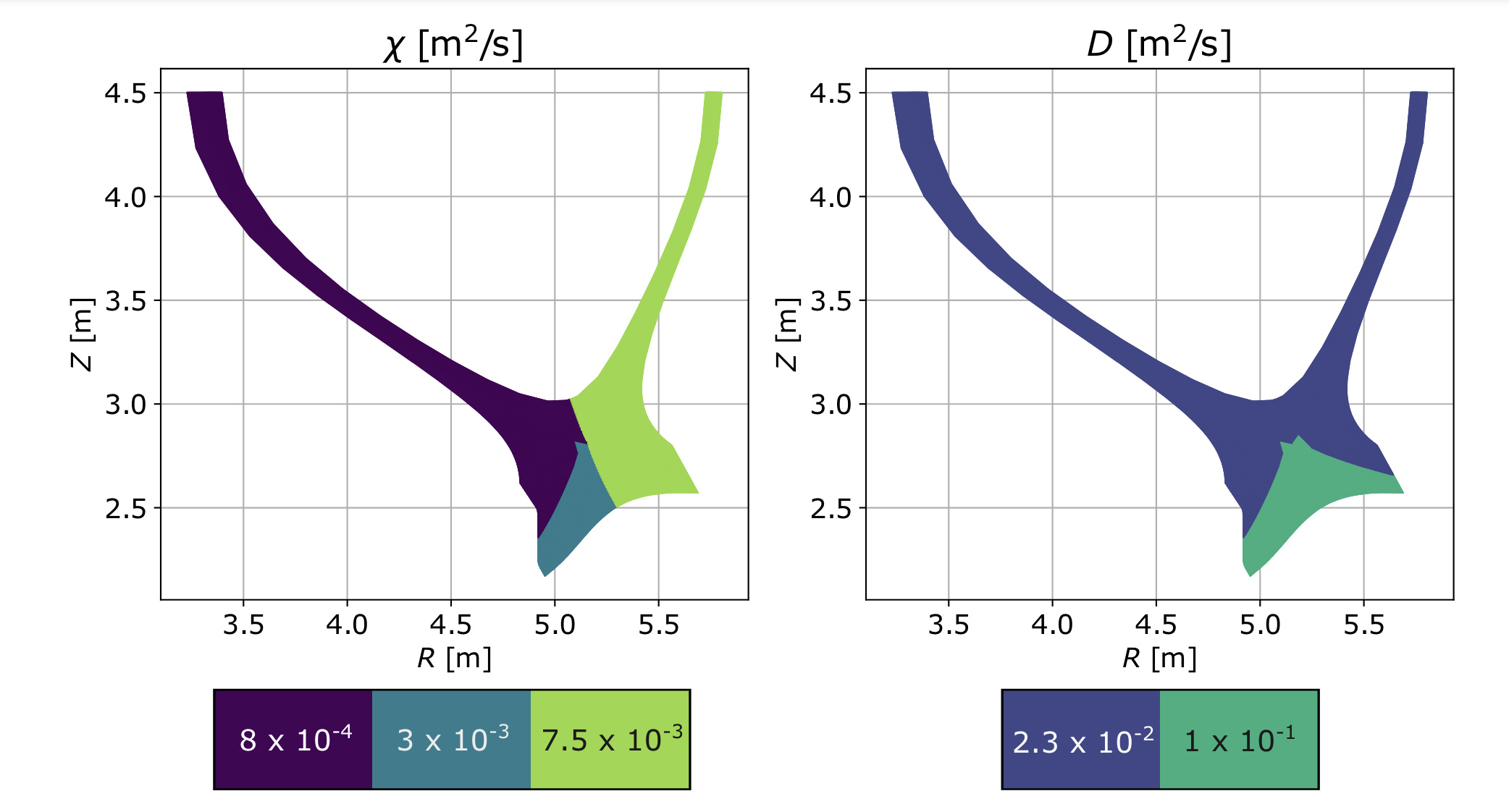}
\caption{2D spatial distributions of $\chi$ (=$\chi_{e}$ = $\chi_{i}$) and $D$. $D$ is constant across the common flux region and in the core, while $\chi$ is increased in the outboard half of the domain. D is increased in the entire PFR, while $\chi$ is increased only in the inboard PFR.}
\label{fig:transport_contours}
\end{figure}

In order to reach sub-mm values for $\lambda_{q}$ consistent with the $B_{p}$ and $<p>$ scalings, $\chi_{e,i}$ had to be reduced to below 10$^{-2}$. It was found that a value of $\chi_{e,i} = 7.5 \times 10^{-3}$ m$^{2}$s$^{-1}$ was sufficient to achieve $\lambda_{q} \approx 0.76$ mm at the entrance to the outer divertor. In order to yield larger density gradient scale lengths, the particle diffusivity was slightly increased, relative the thermal diffusivities. $D = 2.4 \times 10^{-2}$ m$^{2}$s$^{-1}$ yielded a $\lambda_{n}$ at the OMP separatrix of 7.1 mm. $D$ also influences $\lambda_{q}$ through the convective term in the heat equation, $\propto D\nabla n T$, so $D$ was first used to match $\lambda_{n}$ and then $\chi_{e,i}$ was adjusted to yield the desired $\lambda_{q}$. Finally, in order to replicate the expected in-out power asymmetry, poloidal variation was introduced in $\chi_{i,e}$. The grid was split along a radial cut connecting the center of the plasma to the X-point, and $\chi_{i,e}$ on the inboard side was dropped further, to $8 \times 10^{-4}$ m$^{2}$s$^{-1}$. These transport coefficients were applied in the common flux region (CFR) of both the inboard and outboard side individually. It was found that for numerical stability reasons, the transport coefficients had to be increased in the PFR. Plasma in this region is generally cold and rather tenuous, so differences in transport in this region are not expected to have a large impact on the overall solution. Figure \ref{fig:transport_contours} shows the spatial variation of transport coefficients required to reproduce the gradient scale lengths outlined in this section.

\begin{table}[h]
\begin{center}
\caption{Principal control parameters used in UEDGE simulation.}
\label{tab:uedge_params}
\begin{tabular}{lll}
\hline
Parameter & Value & Units \\
\hline
\hline
$P_\mathrm{SOL}$ & 25 & MW \\
$n_\mathrm{sep}$ & 0.96 & $\mathrm{10^{20} m^{-3}}$ \\
$\Gamma_{\rm pump}$ & 5 $\times 10^{21}$ & $\mathrm{s^{-1}}$ \\
\hline
\end{tabular}
\end{center}
\end{table}

\subsection{Results of base case}
\begin{figure}
\centering
\includegraphics[width=1\columnwidth]{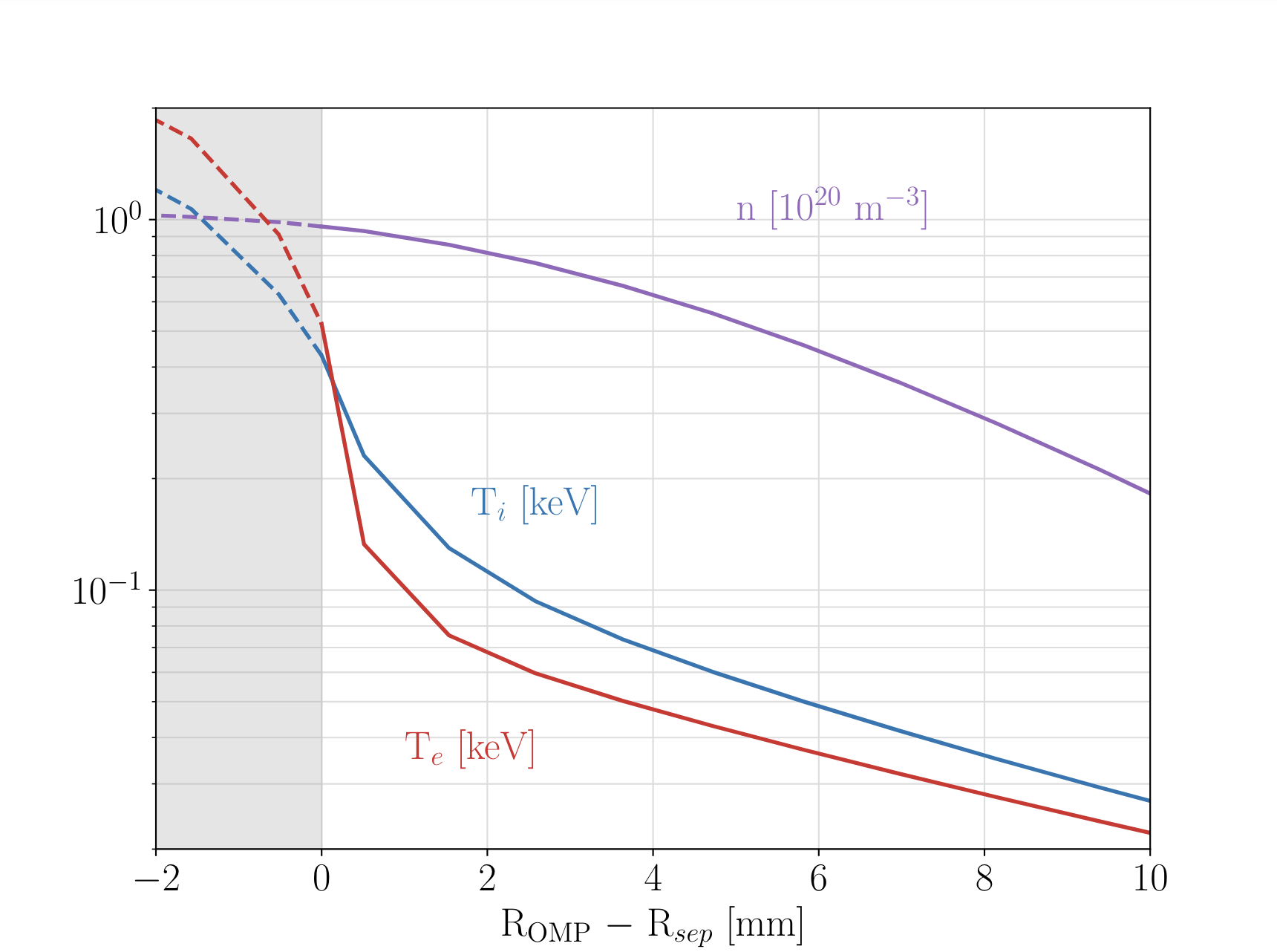}
\caption{Upstream profiles of $n$ (purple), $T_{i}$ (blue), and $T_{e}$ (red) at the OMP. The density gradient scale lengths at the separatrix are larger than temperature gradient scale lengths.}
\label{fig:uedge_upstream}
\end{figure}

\begin{figure*}[hbtp!]
\centering
\includegraphics[width=1.8\columnwidth]{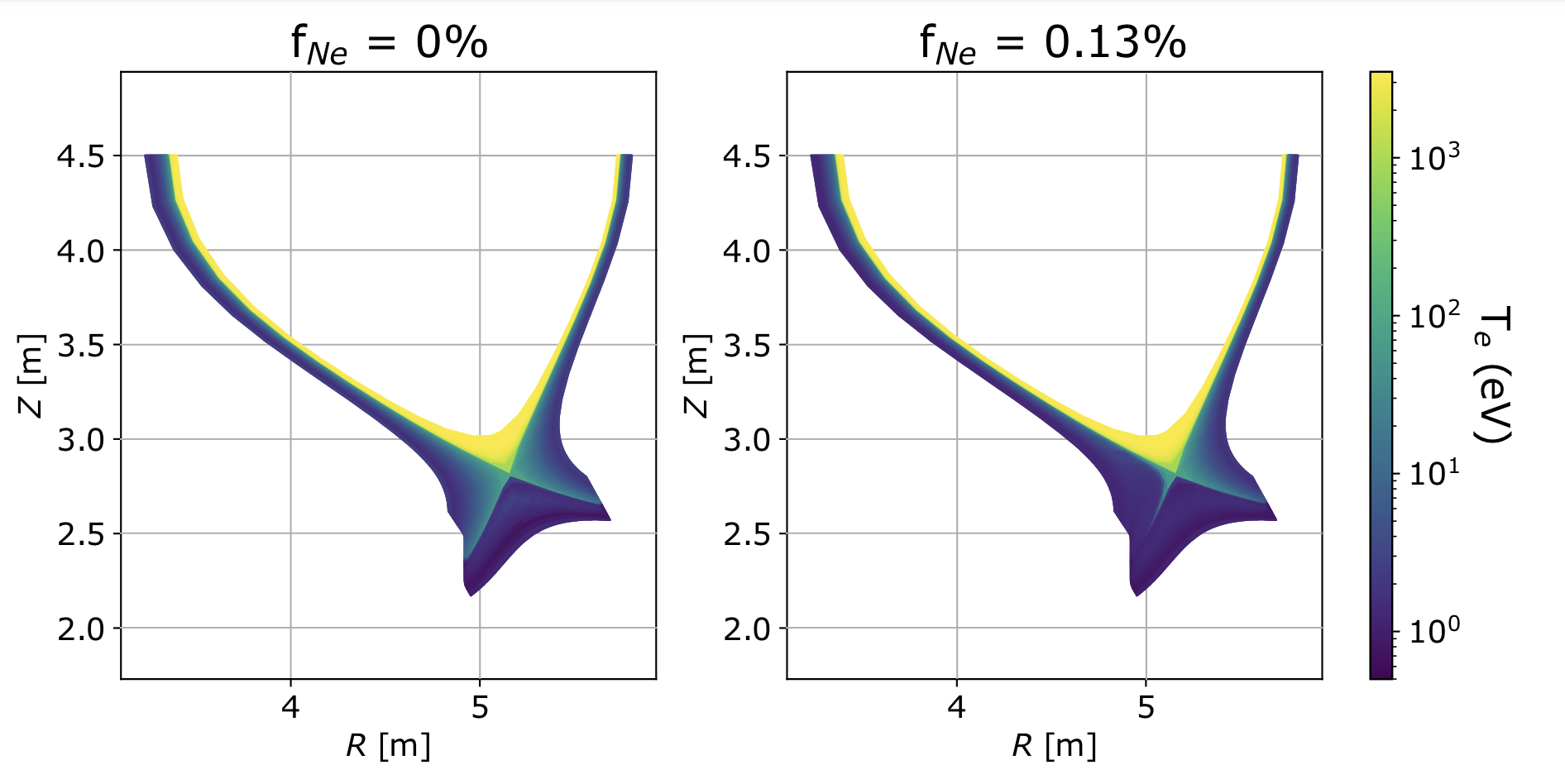}
\caption{2D contours of electron temperature ($T_{e}$) of UEDGE solutions for both no external impurity seeding, $f_{Ne}$ = 0\% (left) and some impurity seeding, $f_{Ne}$ = 0.13\% (right).}
\label{fig:uedge_soln}
\end{figure*}

\begin{figure*}[h!]
\centering
\includegraphics[width=1.8\columnwidth]{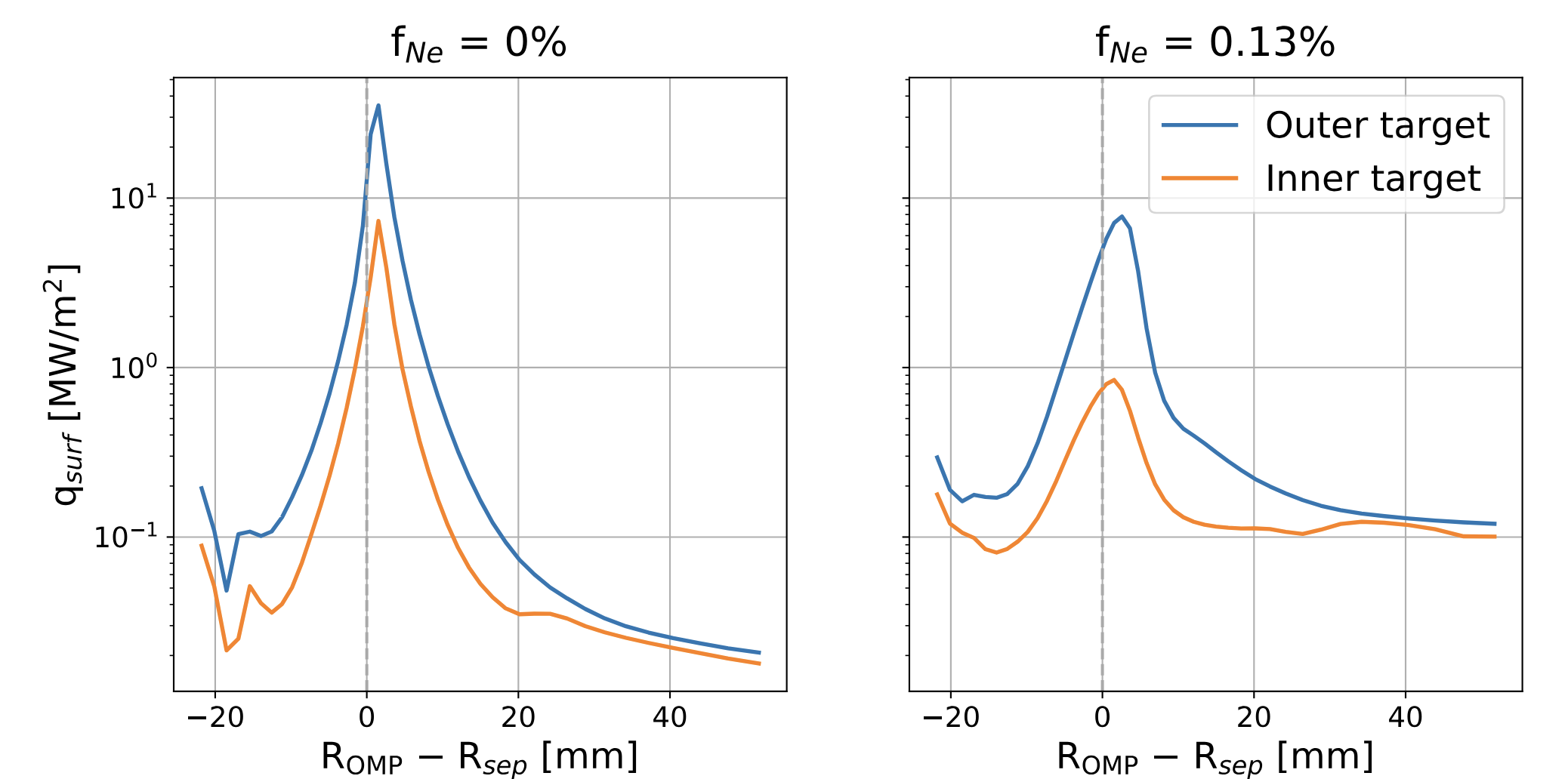}
\caption{Total heat flux density arriving at the inner (orange) and outer (blue) targets in both unseeded, $f_{Ne}$ = 0\% (left), and seeded, $f_{Ne}$ = 0.13\% (right), cases.}
\label{fig:uedge_targets}
\end{figure*}

Table \ref{tab:uedge_params} records the three primary plasma parameters used as inputs for UEDGE, discussed in Sections \ref{subsec:sources} and \ref{subsec:transport}. Figure \ref{fig:uedge_upstream} shows the resulting upstream profiles for these parameters as well as the transport coefficients shown in Figure \ref{fig:transport_contours}. The plasma density, $n$, clearly decays more slowly into the SOL than $T_{e}$ and $T_{i}$, which have small $\lambda_{T}$ consistent with the small $\lambda_{q}$ expected from high $B_{p}$. $T_{i}$ is higher than $T_{e}$ across most of the SOL. At the separatrix, $T_{i}$ falls off more slowly than $T_{e}$, but the gradient scale lengths resemble each other more closely in the far-SOL. The left panel of Figure \ref{fig:uedge_soln} shows the 2D distribution of $T_e$, with a hot core plasma inside the LCFS that cools towards the walls and divertor targets. Regardless, the parallel temperature gradient is weak, an indication of an attached plasma with little power dissipation. In-out differences in $T_{e}$ are not obvious without impurity seeding, but much larger temperatures remain in the outer SOL than in the inner SOL for the seeded case, consistent with detachment beginning in the inner target before the outer. The left panel of Figure \ref{fig:uedge_targets} shows a very peaked heat flux density profile both at the outer and inner targets. This is also characteristic of an attached divertor, in which the heat flux on the target is dominated by the thermal plasma component. The peak heat flux density for this plasma is $q_\mathrm{surf} \approx$ 35 MW/m$^{2}$, significantly above the limit for the divertor's material tolerances.

\subsection{Results of dissipative base case used as MANTA point design}
To dissipate some of this power, MANTA's divertor uses impurity seeding, a technique routinely used for removing power to the targets via impurity radiation \cite{loarte_plasma_1998, leonard_scaling_2012, kallenbach_impurity_2013}. This technique is the same used in MANTA's core to achieve a high radiated fraction. To model extrinsic impurity injection, a fixed-fraction impurity model is used in UEDGE, with neon (Ne) as the radiative species. This model sets the neon fraction of the main ion density, $f_\mathrm{Ne} = n_\mathrm{Ne}/n$, where $n_\mathrm{Ne}$ is the density of Ne. The value for $f_\mathrm{Ne}$ was increased until enough power was radiated in the divertor to reduce $q_\mathrm{surf}$ within material tolerances. It was found that for $f_\mathrm{Ne}$ = 0.13\%, $q_\mathrm{surf}$ on the more heavily loaded outer target dropped well below 10 MW/m$^{2}$.

The right panels of Figures \ref{fig:uedge_soln} and \ref{fig:uedge_targets} show the effects of additional impurity injection on the plasma most clearly. As the impurity fraction is increased, the plasma at the divertor targets, in particular at the inner target, gets colder. The heat flux density at the inner target broadens and drops to below 1 MW/m$^{2}$, indicative of detachment. This is largely a result of the lower power sharing to the inner divertor, driven by reduced turbulence in the inner SOL, and facilitating detachment at that target. Regardless, Figure \ref{fig:uedge_targets} shows that the injection of impurities does also affect the outer divertor leg. The peak heat flux density on the outer target drops more than four-fold, to a more manageable value, $q_\mathrm{surf}$ = 7.8 MW/m$^{2}$. 

\begin{figure}[hbtp!]
\centering
\includegraphics[width=\columnwidth]{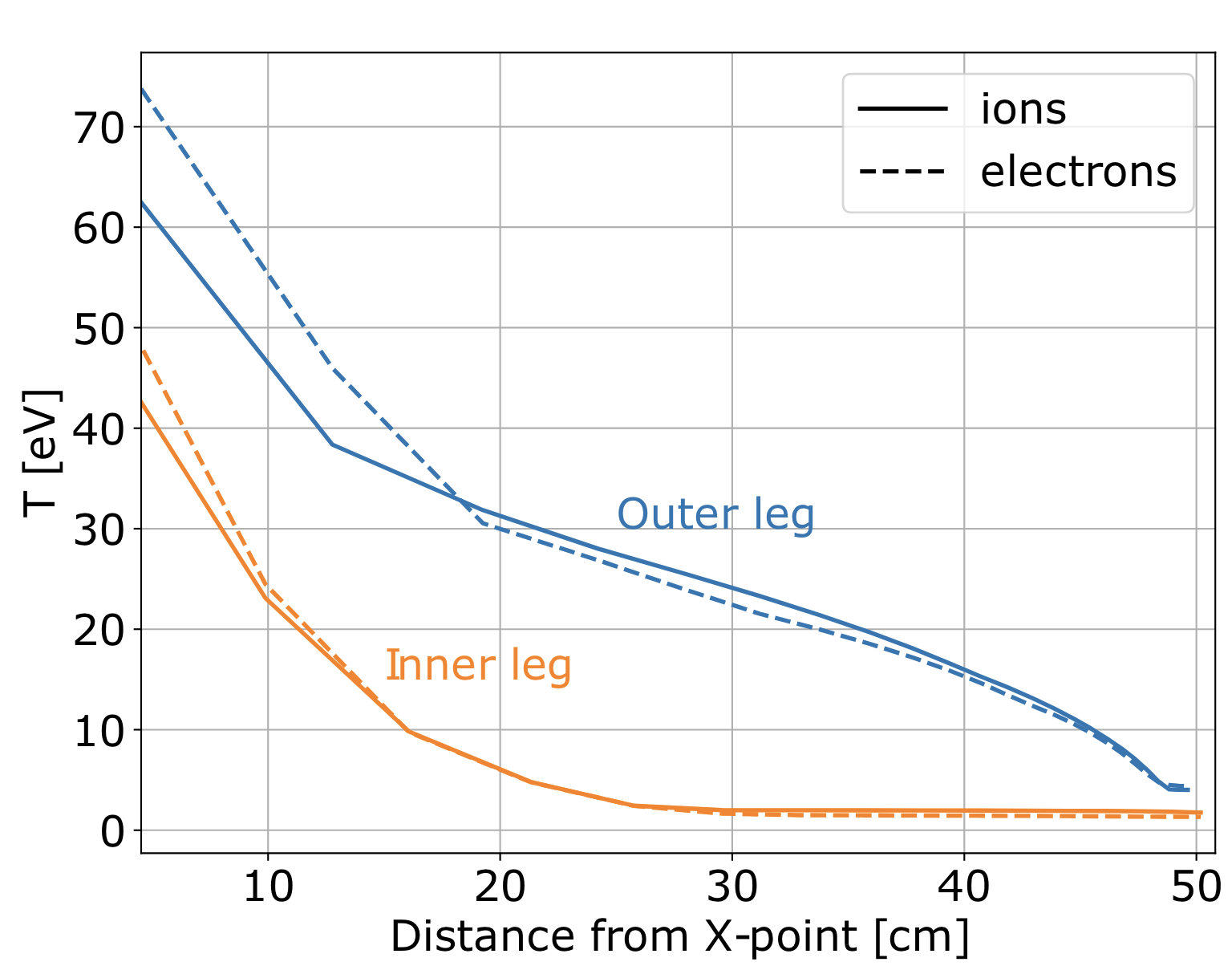}
\caption{Profiles of maximum temperature along outer (blue) and inner (orange) divertor legs, for both ions (solid) and electrons (dashed). Temperatures drop and flatten more quickly along the inner leg and begin to flatten only near the target ($>$ 45 cm along the outer leg).}
\label{fig:par_temps}
\end{figure}

\begin{figure}[hbtp!]
\centering
\includegraphics[width=1\columnwidth]{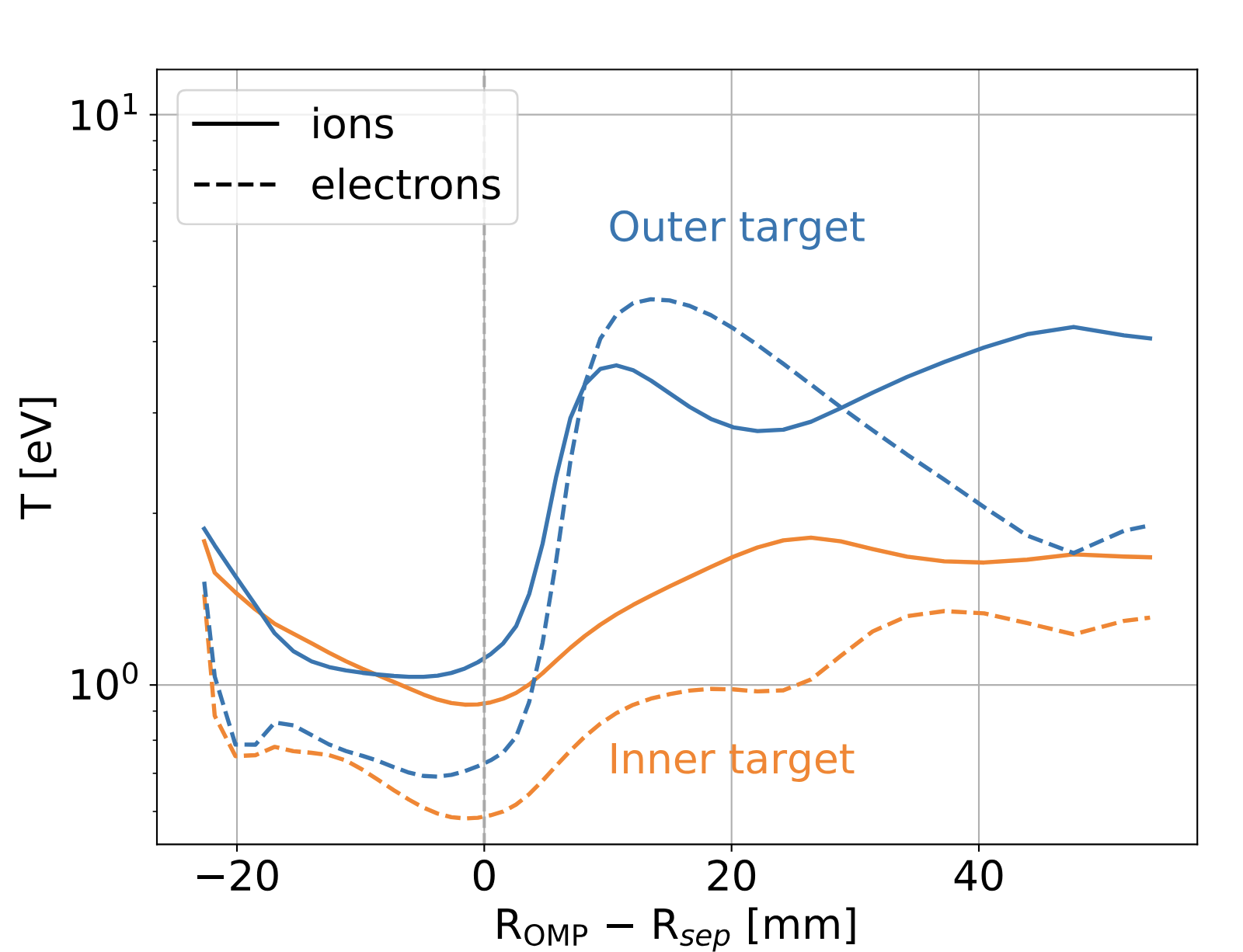}
\caption{Temperatures along inner (orange) and outer (blue) targets for both ions (solid) and electrons (dashed). Temperatures near the strike point are well below 5 eV on both targets. $T_e$ and $T_i$ peak $\sim$10-15 mm from the strike point.}
\label{fig:target_temps}
\end{figure}

Figures \ref{fig:par_temps} and \ref{fig:target_temps} highlight some of the details of this power dissipation. Figure \ref{fig:par_temps} shows the maximum temperatures, both $T_{e}$ and $T_{i}$ along each of the divertor legs. Not unsurprisingly, temperatures on the inner leg are lower and decay to within 10 eV at a distance of just over 15 cm from the X-point. At the same distance, temperatures along the outer leg remain above 30 eV. Local flattening of the parallel temperature gradient only occurs right in front of the target ($\sim$49 cm from the X-point along the outer leg). While some difference between $T_{e}$ and $T_{i}$ exists close to the X-point, halfway to the target, temperatures have equilibrated, likely a result of increased collisions as a result of larger plasma density. By the time plasma arrives at each of the targets, however, it is below 5 eV, the typically assumed limit for target sputtering \cite{brooks_analysis_2013}. Figure \ref{fig:target_temps} shows target temperature profiles. At the inner target, both $T_{e}$ and $T_{i}$ are relatively flat, while at the outer target, some structure remains in the profiles. The peak temperature is $\sim$10 - 15 mm from the strike point, indicating a significant reduction in heat flux resulting from plasma convection.

\section{Operational space of MANTA's divertor}
\label{sec:operational_space}

In the interest of generalizing these results to a broader divertor operational space for MANTA-like NT reactors, scans about the design point presented in previous sections are performed. In particular, these scans explore the sensitivity of $q_\mathrm{surf}$ to the input control parameters. This approach also helps evaluate the robustness of the point solution and accounts for some of the uncertainty in the model parameters and the model itself. To this end, 2D scans of the three most important parameters in mediating heat fluxes on the divertor targets are performed. The parameters chosen are $P_\mathrm{SOL}$, $f_\mathrm{Ne}$, and $n_\mathrm{sep}$. Two sets of scans are performed. The first fixes $f_\mathrm{Ne}$, and scans $P_\mathrm{SOL}$ and $n_\mathrm{sep}$. The second fixes $n_\mathrm{sep}$ and scans $P_\mathrm{SOL}$ and $f_\mathrm{Ne}$. This is done for two different values $n_{sep}$, one at ~0.36$n_{G}$ and one at ~0.5$n_{G}$, one below and one above the nominal operating $n_\mathrm{sep}$.

Each point in the scan represents a converged UEDGE solution. For the transport model chosen in Section \ref{subsec:transport}, across the explored paremeter space, $q_\mathrm{surf}$ at the outer target, $q_\mathrm{surf}^\mathrm{OT}$ is larger than that at the inner target, $q_\mathrm{surf}^\mathrm{IT}$ (i.e. $q_\mathrm{surf}^\mathrm{OT}< q_\mathrm{surf}^\mathrm{IT}$). Thus, only $q_\mathrm{surf}^\mathrm{OT}$ is reported, depicted as the color dimension in the plots in this section. Each of these plots also shows the $q_\mathrm{surf}$ = 10 MW/m$^{2}$ curve, linearly interpolated using this color variable and the abscissa of each plot. Where no partially detached solutions exist, the black dotted line is constructed via extrapolation.

The scans are restricted to the solution space in which the divertor remains in a state of detachment similar to that of the main operating point - that is, the inner legs detach, while the outer legs remain partially attached. This is a practical limit, due to difficulty in achieving stable convergence when transitioning between attached/detached target solutions. Higher $n_\mathrm{sep}$, lower $P_\mathrm{SOL}$, and higher $f_\mathrm{Ne}$ detaches the outer legs. The opposite reattaches the inner legs. For simplicity, $D$ and $\chi$ are kept constant in these scans. 

\subsection{Scan of upstream density and power at fixed impurity fraction}

The first scan takes the impurity fraction of neon to be fixed at $f_{Ne} = 0.35\%$. This impurity fraction is somewhat higher than that required to just barely meet divertor tolerances as shown in Section \ref{sec:vv_results}, which provides considerable margin for the basecase input parameters parameters listed in Table \ref{tab:uedge_params}. With this fixed, scans in $n_\mathrm{sep}$ and $P_\mathrm{SOL}$ are performed. While neither of these can be controlled with absolute certainty, they represent fairly good proxies for two of the main control room actuators, fueling and heating, respectively. The scan in density is particularly useful because uncertainty in predictive capabilities for $n_\mathrm{sep}$ given a particular fueling scheme renders it vital to scan about a value extrapolated from current experiments. On the other hand, even if predictions for the point value of $n_\mathrm{sep}$ are correct, it may still be desirable from a performance point of view to increase the density in an attempt to access larger fusion gain. Core modeling of MANTA shows that for a fixed value of $P_\mathrm{SOL}$, MANTA can generate more fusion power simply by increasing the line-averaged density,  $\bar{n}$, \cite{rutherford_manta_2024}, which might also raise $n_\mathrm{sep}$. Varying the density also opens the possibility of a larger upper limit for $P_\mathrm{SOL}$ because of the ability to dissipate more power at higher $n_\mathrm{sep}$. Furthermore, heating in a burning plasma will be dominated by $P_{\alpha}$, and so fueling is expected to be an even more important actuator in a reactor than $P_\mathrm{aux}$. Higher $n_\mathrm{sep}$ may then offer a higher margin for error for $P_\mathrm{SOL}$, inspiring more confidence in an integrated core-edge solution.

\begin{figure}[h]
\centering
\includegraphics[width=0.9\columnwidth]{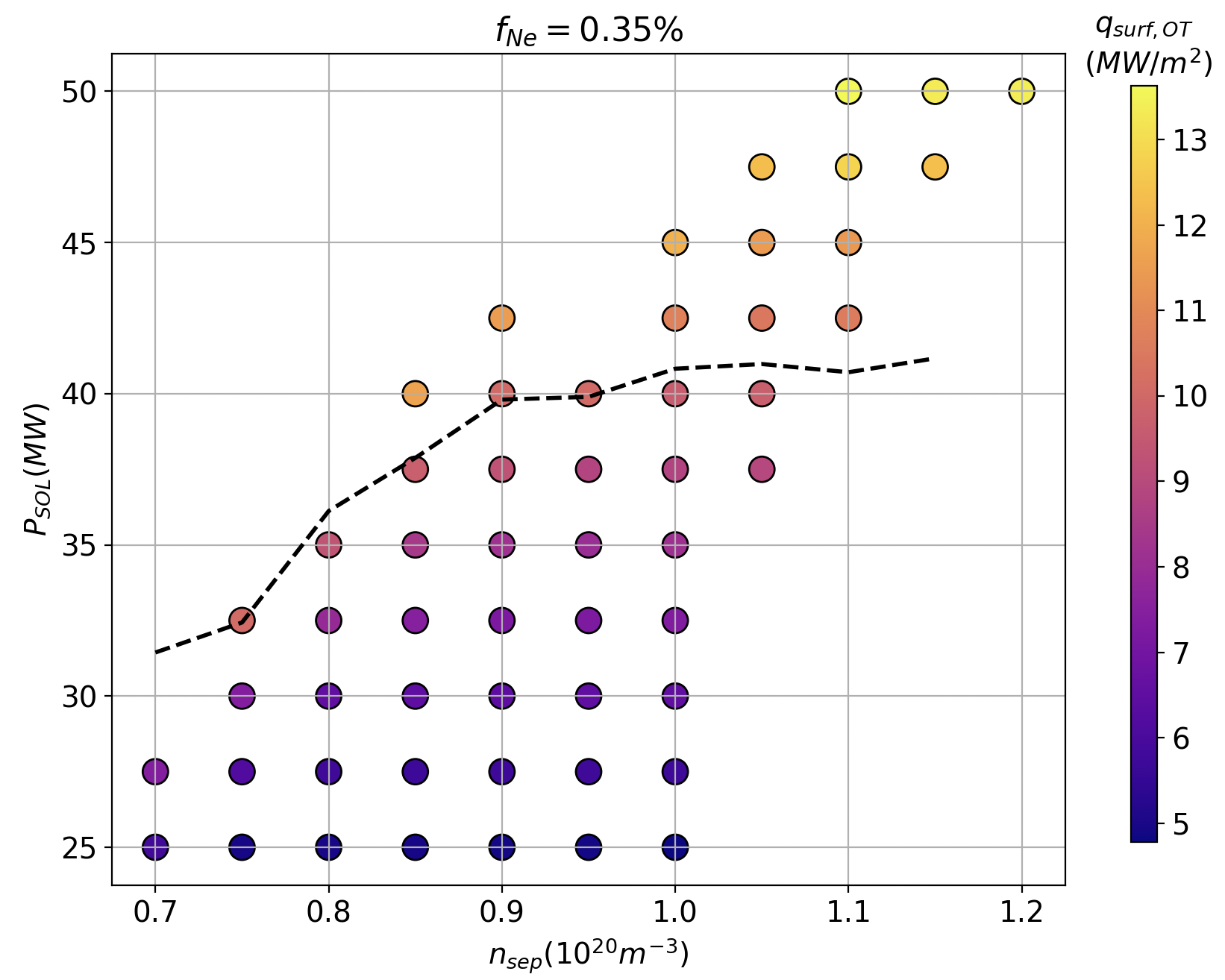}
\caption{2D scan of $n_\mathrm{sep}$ and $P_\mathrm{SOL}$ at fixed $f_\mathrm{Ne} = 0.35\%$. The color bar shows $q_\mathrm{surf}$ at the outer target surface at the different points in the operational space. The black dashed curve demonstrates the boundary below which the $q_\mathrm{surf,OT} >$ 10 MW/m$^{2}$. Only points that converged and kept the inner leg detached and the outer leg attached are shown.}
\label{fig:nsep_scan}
\end{figure}

To vary $n_\mathrm{sep}$, the density boundary condition at the innermost radial flux surface is varied. At differing values of this boundary condition, the power entering the simulation domain at this same boundary is varied. Robust divertor operation is found for $P_\mathrm{SOL} > 25$ MW, so long as $n_\mathrm{sep}$ remains sufficiently high. Figure \ref{fig:nsep_scan} shows a scan in $n_\mathrm{sep}$ from $0.7 - 1.2 \times 10^{20} $m$^{-3}$ and a scan in $P_\mathrm{SOL}$ from 25 - 50 MW. Regardless, it is found that past $\sim$40 MW, increases in $n_\mathrm{sep}$ do not help considerably with power dissipation. Higher values of $P_\mathrm{SOL}$ may require additional impurity seeding, as described below.

\subsection{Scan of power and impurity fraction at fixed upstream density}

\begin{figure*}[hbtp!]
\centering
\includegraphics[width=2\columnwidth]{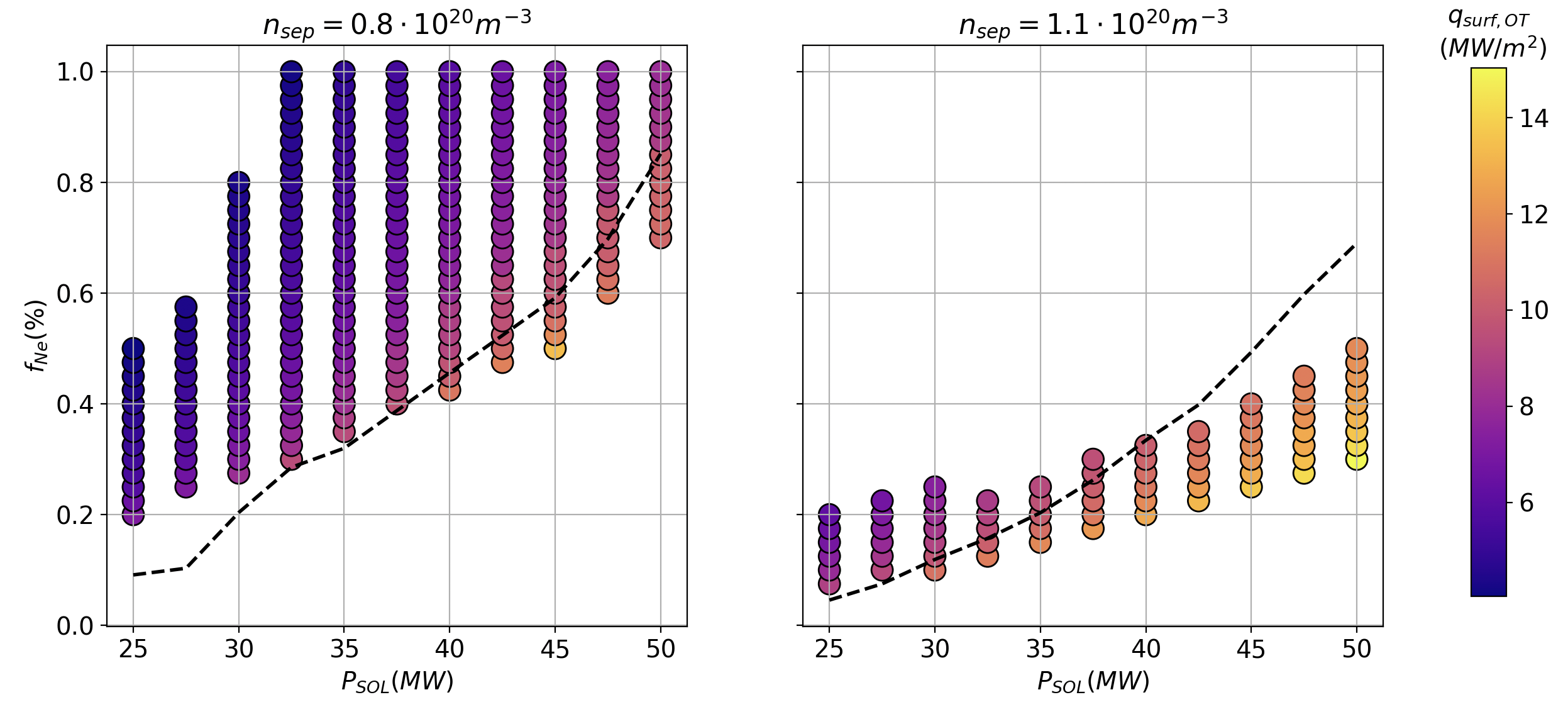}
\caption{2D scan of $P_\mathrm{net}$ and $f_\mathrm{Ne}$ at fixed $n_\mathrm{sep} = 0.8 \times 10^{20} $m$^{-3}$ (left) and $n_\mathrm{sep} = 1.1 \times 10^{20} $m$^{-3}$ (right). These two points represent slightly higher and slightly lower densities than the nominal operating $n_\mathrm{sep} = 0.96 \times 10^{20} $m$^{-3}$. The color bar shows the heat flux at the outer target surface at the different points in the operational space. The black dashed curve demonstrates the boundary below which the $q_\mathrm{surf,OT} >$ 10 MW/m$^{2}$.}
\label{fig:psol_scan}
\end{figure*}

A second set of scans explores how MANTA's operators may need to adjust the level of injected neon in order to maintain acceptable heat fluxes in the divertor in the event that the upstream value for $P_\mathrm{SOL}$ varies. Or conversely, it may inform leniency in the allowable level of $P_\mathrm{SOL}$ if MANTA can reliably regulate the Ne content in its divertor. Since $n_\mathrm{sep}$ remains uncertain, scans are performed at both a lower density and a higher density, $n_\mathrm{sep} = 0.8 \times 10^{20}$m$^{-3}$ and $n_\mathrm{sep} = 1.1 \times 10^{20}$m$^{-3}$. These correspond to separatrix Greenwald fractions, $n_\mathrm{sep}/n_{G}$, of 0.36 and 0.5 respectively. Taking $\bar{n} = 2n_\mathrm{sep}$ would give overall Greenwald fractions for these two operating points of $f_{G} \approx 0.72$ and $f_{G} \approx 1$ respectively. Of course $f_{G}$ at unity is nominally an unstable operating point. Recent results in DIII-D's NT campaign, however, show operation even at $f_{G} \approx 2$, even though its edge remained below $n_{G}$, implying greater density peaking than the assumption made here of $\bar{n} = 2n_\mathrm{sep}$. If MANTA can achieve this level of peaking, it is conceivable that MANTA may also be able to operate at this density. 

This high density scan presents a potential, if optimistic, expansion to MANTA's operational space if robust high density operation in NT is successful. 

These scans vary $P_\mathrm{SOL}$ between $25 - 50$ MW and $f_\mathrm{Ne}$ between $0 - 1\%$. As with the previous set, they demonstrate a very flexible operational space for MANTA's divertor. The left panel of Figure \ref{fig:psol_scan} shows that even at low density, it remains within 10 MW/m$^2$ for $P_\mathrm{SOL}$ up to 50 MW with $f_\mathrm{Ne} <  1\%$. The $q_\mathrm{surf}$ line is linear throughout most of the scan and perhaps becomes quadratic only at high $P_\mathrm{SOL}$. It is a good rule of thumb then, that the amount of additional impurity needed should scale as the amount of additional power crossing the separatrix.

At  high density, shown in the panel on the right of Figure \ref{fig:psol_scan} the operational space is somewhat reduced. This is a result of the imposition that the state of detachment remain the same as in the standard operating scenario, i.e. inner leg detached, outer leg attached. Whether or not this might be an actual requirement of this plasma or is a physical result is outside the scope of this paper. In some sense, the fact that partially detached solutions exist only at fairly moderate impurity fractions is a testament to the relative ease with which an NT plasma may detach. If full detachment does not compromise core performance, then this regime might be easily reached. Regardless, even with the requirement of partial detachment it is found that very little impurity seeding is required. At $P_\mathrm{net}$ up to $\sim$35 MW, $f_\mathrm{Ne} < 0.2$ is sufficient to keep $q_\mathrm{surf} < 10$ MW/m$^2$.

\section{Divertor and vacuum vessel design}
\label{sec:vv_design}

\begin{figure*}
\centering
\includegraphics[width=1.8\columnwidth]{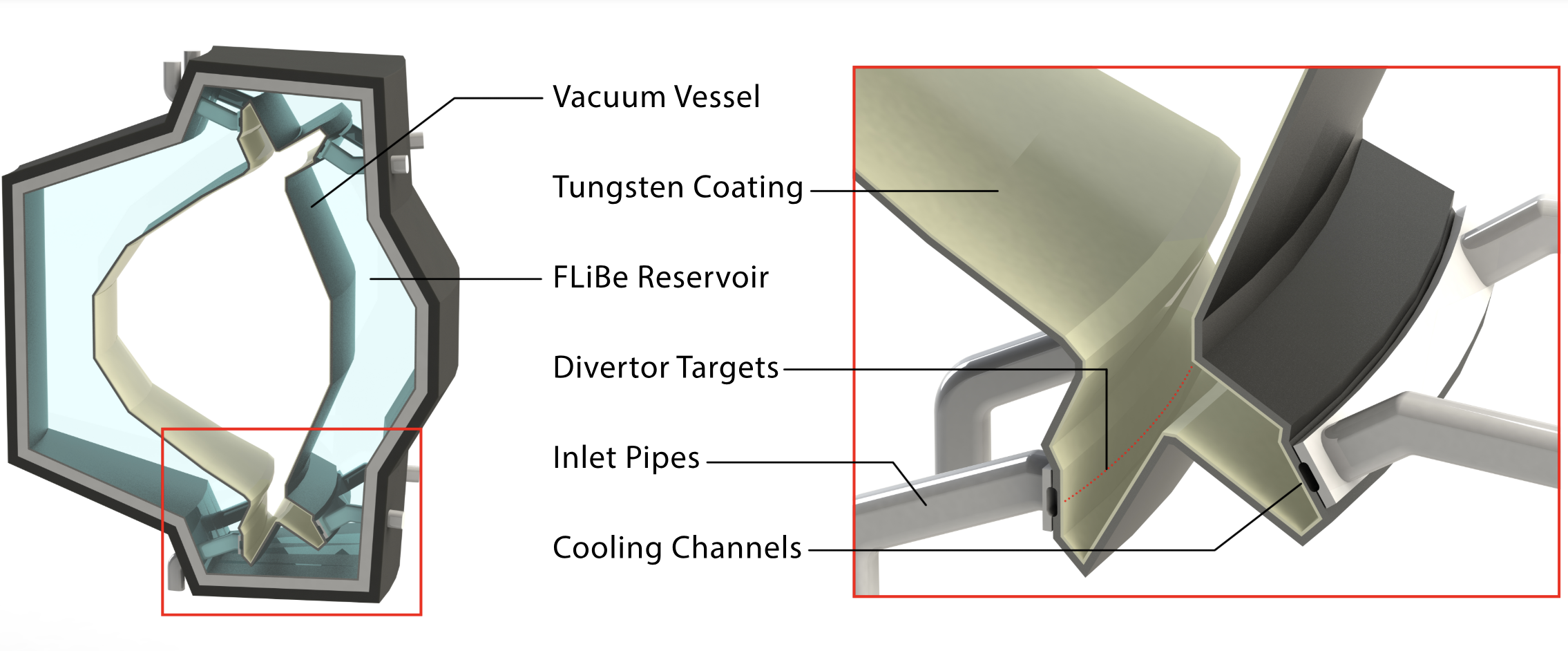}
\caption{CAD models for vacuum vessel and FLiBe reservoir (left) and blow-up of divertor-specific components (right). Inlet pipes circulate FLiBe along the divertor targets in the toroidal direction, through cooling channels directly in contact behind the vacuum vessel onto which a tungsten coating is added.}
\label{fig:Vacuum Vessel and Divertor Cross Section}
\end{figure*}

With both a solution for a highly dissipative plasma edge and scans about this solution demonstrating how different values for $q_\mathrm{surf}$ might be achieved, it was then possible to design the material surfaces of MANTA's plasma-facing components (PFC). The FreeGS equilibrium was used as a template on which to define the interior boundary of the vacuum vessel. The resulting contour is shown in gold in Figure \ref{fig:Vacuum Vessel and Divertor Cross Section}. Note that disruption loads are not considered here, but would add additional structural requirements. This contour then informed decisions about dimensions of the tungsten coating, VV wall, and FLiBe cooling channels, all of which lie recessed behind the inner vacuum vessel boundary. Using an estimate from core modeling for the amount radiated power that would arrive at the VV wall, Ansys Fluent thermal conduction calculations found that the VV wall would require active cooling. 

\begin{figure}
\centering
\includegraphics[width=\columnwidth]{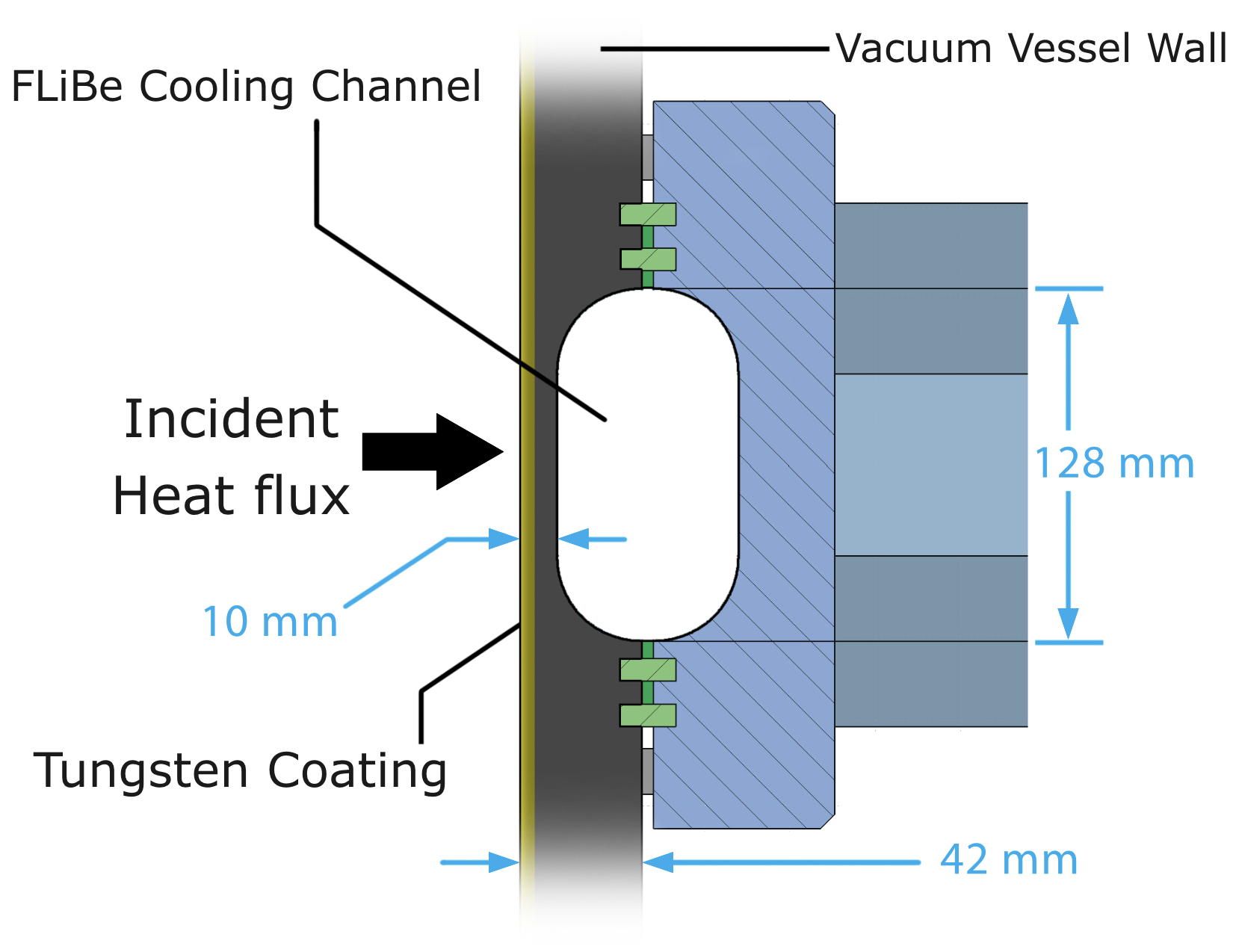}
\caption{Poloidal cross section of divertor target cooling interface design. Plasma fluxes enter from left of the figure and come into contact with a tungsten coating on top of the VV wall. A cooling channel carrying FLiBe sitting right of the VV wall dissipates the incident heat flux. The tungsten plasma-facing surface is shown in gold, the alloy sealing gaskets in green, and the external channel block in blue. Figure reproduced from \cite{rutherford_manta_2024}.}
\label{fig:divertorPanelCrossSection}
\end{figure}

Most of the heat load, however, arrives at the divertor targets, and effective power exhaust therefore required FLiBe cooling channels in direct contact with the strike point. To minimize complexity the divertor is divided into 18 identical $20 \degree$ segments, matching the spacing of the TF coils. Each segment consists of a long groove in the VV exterior, which becomes a cooling channel when an exterior block with a matching groove seals the channel shut. The panels follow a smooth $20 \degree$ arc with a roughly 5 m radius, which results in a panel approximately 2 m long. The FLiBe cooling channel features double-wall gaskets to accommodate the pressure and corrosion considerations of FLiBe. An example of the outboard divertor segment design can be seen on the right hand side of Figure \ref{fig:Vacuum Vessel and Divertor Cross Section}. A detailed depiction of the divertor cooling channel interface, including dimensions, is given in Figure \ref{fig:divertorPanelCrossSection}. The thin layer between the incident heat flux and the cooling channel is favorable from the perspective of stationary power handling, but would also need to be tested against transient loads, such as disruptions. 

\section{Heat transfer simulations of divertor targets}
\label{sec:vv_results}

\begin{figure}
\centering
\includegraphics[width=\columnwidth]{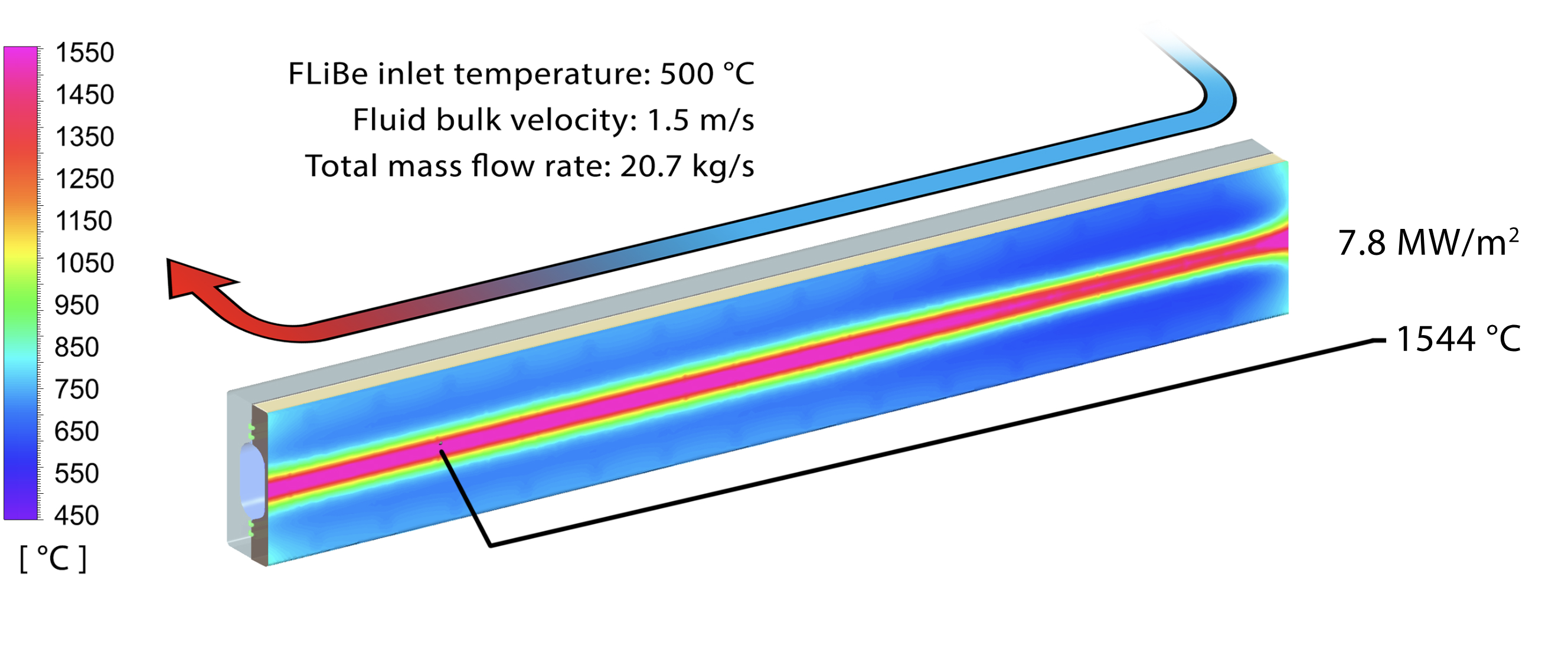}
\caption{2D temperature distribution on divertor outer target resulting from an Ansys Fluent heat transfer simulation. As the FLiBe heats up from contact with the panel, the bulk temperature along the panel increases. Regardless, for a peak heat flux density of 7.8 MW/m$^{2}$, the maximum temperature remains just under the recrystallization temperature of tungsten ($1550\degree$C) \cite{deformedTung}. Details of coolant flow in the simulations are shown as well.}
\label{fig:comparison_surf_temps}
\end{figure}

Before finalizing the design of the divertor targets, Ansys Fluent \cite{ANSYS} was used to perform a full heat transfer calculation and ensure the design met the required target cooling specifications. To do this, the heat flux density profile from the more heavily loaded outer divertor in UEDGE was imposed onto the model for the divertor target described in Section \ref{sec:vv_design}. To facilitate the calculation, the heat flux profile from UEDGE was approximated as a Gaussian with a a peak value of 7.8 MW/m$^{2}$ and a fall-off length of 5.6 mm, consistent with the predicted upstream $\lambda_{q} \approx 0.75$ mm and flux expansion near the divertor target. This profile is assumed to be toroidally symmetric on the divertor target.

\begin{figure*}[h!]
\centering
\includegraphics[width=2\columnwidth]{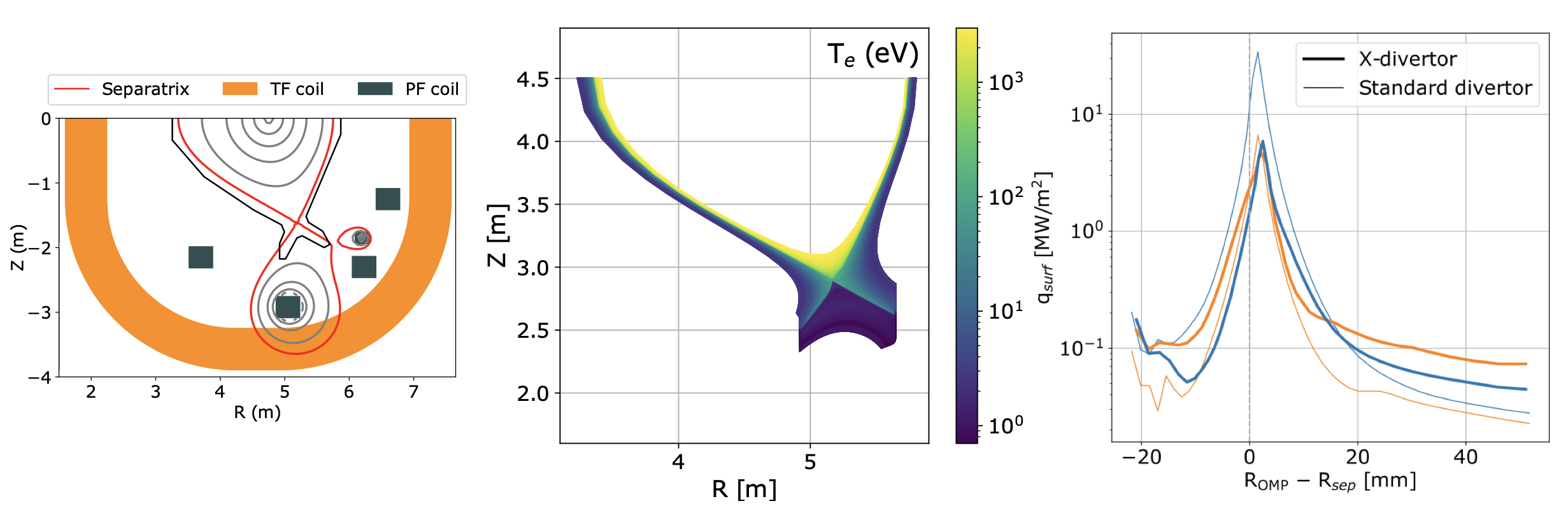}
\caption{From left to right: coil set used to generate the equilibrium required for the X-divertor configuration; $T_{e}$ from UEDGE solution computed on mesh requiring a vertical outer target to ensure numerical stability resulting from strike angles below 1\degree; heat flux profiles on the outer (blue) and inner (orange) targets for $f_{Ne}$ = 0.01\%.}
\label{fig:xdiv}
\end{figure*}

Primary heat removal from the divertor plate is modeled by specifying the inlet velocity and temperature for the FLiBe running through the cooling channel. Using thermal conductivity and heat transfer coefficients for the tungsten and between the cooling interface, Ansys Fluent computes the full 2D temperature distribution on the divertor target. In these simulations, the VV wall was considered a single piece of tungsten to avoid simulation artefacts in the thin geometries. This simplified model also neglected toroidal curvature, fasteners, as well as any additional heat removal through the gasket interface. The resulting target temperatures for this heat flux profile as well as a heat flux profile with about half as high $q_\mathrm{surf}$ are shown is shown in Figure \ref{fig:comparison_surf_temps}. The maximum temperature for the nominal operating point is $\sim$1540$\degree$C, only marginally below the recrystallization temperature of tungsten ($1550\degree$C \cite{deformedTung}). This sets a strict upper limit for the allowable heat flux density on the divertor.

\section{Advanced divertor concepts}
\label{sec:adv_div}

While the goal of MANTA's design was to keep the divertor equilibrium as simple as possible, it is interesting to explore the possibility of integrating advanced divertor concepts. To this end, FreeGS was used to generate a new equilbrium with a second X-point placed directly behind the outer divertor targets - known as an 'X-divertor'. This concept increases flux-expansion near that target, with the intention of spreading power more evenly across the target surface \cite{kotsch_xdiv_2004, covele_x-divertors_nodate}. The X-divertor equilibrium was produced by adding an additional PF coil (PF5) and modifying the locations and currents of the other coils. Despite modifications to the other coils, currents in PF1 - PF4 increase by at most 26$\%$ (note: the current in PF4 changes direction). See Table \ref{tab:X-divertorCoilParams} for relevant parameters of the X-divertor coilset. This new equilibrium was used to generate another computational mesh for UEDGE, following the same guidelines with respect to grid resolution as outlined in Section \ref{sec:uedge}. Using the formula for flux expansion, $f_{x}$, in \cite{anand_plasma_2021} and calculating quantities at the OMP and the outer target separatrix, $f_{x} = 6.0$ for the X-divertor, almost four times as high than for the standard equilibrium outlined in Section \ref{subsec:equ}, which had $f_{x} = 1.6$. The increased flux expansion near the outer target strongly reduces the angle of incidence of the strike point, facilitating the requirement to keep the outer target grazing angle as close to 2$\degree$ as possible without need for additional target tilt. In fact, the outer target needed to be made vertical, as shown in Figure \ref{fig:xdiv}, to keep the tilt above 1$\degree$ in the interest of numerical stability.

\begin{table}[h]
\begin{center}
\caption{X-divertor poloidal field coil parameters}
\label{tab:X-divertorCoilParams}
\resizebox{\columnwidth}{!}{
\begin{tabular}{llllll}
 \hline
 Parameter & PF1 & PF2 & PF3 & PF4 & PF5\\
 \hline
 \hline
Current$\cdot$turns [MA$\cdot$turns] & -7.51& 7.7& 2.3& 5.25& 4.04 \\
Number of turns & 83 & 85 & 26 & 58 & 45\\
Height [m] & 0.33 & 0.33 & 0.18 & 0.29 & 0.25 \\
Width [m] & 0.37 & 0.37 & 0.22 & 0.25 & 0.25\\
R [m] & 3.7 & 5.05 & 6.23 & 6.6 & 6.2 \\
Z [m] & $\pm$2.15 & $\pm$2.92 & $\pm$2.3 & $\pm$ 1.25 & $\pm$ 1.85\\
 \hline
\end{tabular}}
\end{center}
\end{table}

UEDGE was then run on this new mesh, using the same parameters as in Table \ref{tab:uedge_params} and the same target $\lambda_{q}$, $\lambda_{n}$, and $P_\mathrm{in-out}$ as those outlined in Section \ref{subsec:transport}. In order to achieve these, however, some changes to the transport coefficients had to be made. In particular, to achieve the expected power sharing, both $\chi$ \emph{and} $D$ had to be decreased considerably on the inboard side. Even for $D_\mathrm{in} = \chi_\mathrm{in} = 10^{-4}$ m$^{2}$s$^{-1}$ and $\chi_\mathrm{out}$ chosen to match the expected $\lambda_{q}$, sufficiently low values for $P_\mathrm{in-out}$ could not be achieved. As a compromise, $\chi_\mathrm{out}$ was increased slightly to yield $P_\mathrm{in-out}$ closer to the target 0.3, at the cost of a slightly larger $\lambda_{q}$. Whether the difficulty in diverting the experimentally-observed amount of power to the outer target is a numerical difficulty or rather a feature of increased flux expansion in an NT divertor is unclear. It warrants further investigation and could motivate deeper study of alternative divertor concepts in NT shapes.

For this divertor configuration, only 0.01\% Ne fraction was required to bring down target heat fluxes to allowable levels (compared to 0.13\% Ne needed for the standard divertor). In contrast to the standard divertor leg configuration, detachment onset occurred first at the outer leg, rather than the inner leg, likely a result of the increased flux expansion on the outer target. The rightmost panel of figure \ref{fig:xdiv} shows the heat flux profiles on both targets for the X-divertor with 0.01\%, as well as for the standard divertor with the same $f_\mathrm{Ne}$. The inner target only sees a slight reduction in the maximum $q_\mathrm{surf}$, but the difference is marked at the outer target. While more careful equilibrium generation and target geometry design would likely be required to inspire more confidence in the success of this technique, this initial scoping shows promise that advanced divertors could be used in conjunction with standard power mitigation techniques for the already simpler task of dissipating heat in an NT divertor.

\section{Conclusions}
\label{sec:conclusions}

It is clear that the combined effects from ELM-free continuous operation, low $P_\mathrm{SOL}$, and high edge density, are greatly beneficial to a fusion reactor's divertor power exhaust requirements. And beyond metrics involving global quantities, NT divertors benefit from increased target surface area as a result of their location at larger $R$ for fixed device $R_{0}$. Though plasma cross-field transport is notoriously challenging to diagnose and model, results from NT experiments indicate that both the heat flux width and the power sharing between divertor in NT will be significantly more amenable to divertor survivability than that in a conventional PT H-mode reactor at similar $R_{0}$ and $B_{t}$.

To quantify MANTA's divertor challenge beyond these 0D metrics, the first step was to develop the divertor geometry and poloidal field coil set. The magnetic equilibrium generated by these PF coils was then used to construct a simulation grid for the ensuing UEDGE simulations of the boundary plasma dynamics. For the $P_\mathrm{SOL}$ = 25 MW, it was found that only $\sim$0.1\% Neon impurity content was needed to dissipate enough parallel heat flux to ensure the divertor targets received sufficiently low heat fluxes and maintained plasma temperatures below sputtering thresholds. This impurity fraction is significantly lower than that expected to be required for dissipation in PT devices \cite{ballinger_simulation_2021}. Despite the need to actively cool the VV as a result of a high core radiated fraction, it is demonstrated that the same fluid used for tritium breeding in the fuel cycle, FLiBe, can also be circulated in cooling channels behind the PFCs to evacuate the necessary heat from the targets. For a peak heat flux of 7.8 MW/m$^{2}$, the peak temperature at the targets remains below 1550$\degree$C.

Perturbation about the design point at higher and lower densities allow a great deal of flexibility in the operation of MANTA. In particular, they help ease any concerns related to uncertainties in the behavior of an NT core at reactor parameters, as well as known uncertainties in cross-field transport magnitudes and their poloidal distribution. If the avoidance of ELMs in NT remains robust \cite{Nelson_2023, nelson2022prospects, nelson_characterization_2024, saarelma_ballooning_2021, Coda_2022, marinoni_brief_2021}, the simple divertor can handle up to double the standard operating $P_\mathrm{SOL}$ with impurity fractions of only 1\%, although a stricter limit based on core contamination may exist. 

The current work also motivates the ability to further manage power flows to a particular target via flux expansion by placing an additional null behind the divertor target, with only one additional external PF coil. There exists some amount of uncertainty in the scalability of an NT concept to a reactor-class tokamak. A great deal of experimental and theoretical work, largely a result of NT campaigns both at DIII-D and TCV, inspire confidence that NT may be a viable reactor scenario. This work adds to a growing body of results, many experimental, suggesting that an NT divertor will see a less challenging power handling environment, which may provide the necessary margin for testing NT at reactor-relevant parameters and operating scenarios.

\section*{Contributions}
M.A. Miller led writing and UEDGE simulations; D. Arnold worked on UEDGE base case and target plate / VV design; M. Wigram provided direct mentoring and facilitated use of UEDGE; A.O. Nelson provided direct mentoring and background on advances in NT; J. Witham led VV and PFC design and Ansys simulations; G. Rutherford generated FreeGS equilibria and PF coil design; H. Choudhury assisted with UEDGE simulations; C. Cummings assisted with Ansys simulations; C. Paz-Soldan and D.G. Whyte oversaw work.

\section*{Acknowledgements}
The authors would like to acknowledge and thank all other members of courses MIT 22.63/CU-APPH 9143. They would like to also thank Sean Ballinger for use of his UEDGE post-processing tools. This work was supported in part by US DOE grants DE-FC02-04ER54698, DE-SC0014264, DE-SC0022270, DE-SC0022272,  DE-SC0021622, DE-SC0023289, DE-SC0021629, DE-SC0021657, DE-SC0021325, and DE-FG02-86ER53222. Disclaimer: This report was prepared as an account of work in part sponsored by an agency of the United States Government. Neither the United States Government nor any agency thereof, nor any of their employees, makes any warranty, express or implied, or assumes any legal liability or responsibility for the accuracy, completeness, or usefulness of any information, apparatus, product, or process disclosed, or represents that its use would not infringe privately owned rights. Reference herein to any specific commercial product, process, or service by trade name, trademark, manufacturer, or otherwise does not necessarily constitute or imply its endorsement, recommendation, or favoring by the United States Government or any agency thereof. The views and opinions of authors expressed herein do not necessarily state or reflect those of the United States Government or any agency thereof.

\section*{References}
\bibliographystyle{unsrt}
\bibliography{references}

\end{document}